\documentclass[aip,reprint,twocolumn]{revtex4-1}

\usepackage{graphicx}
\usepackage{amsmath}
\usepackage{bbm}
\newcommand{\wherebibJ}{/home/jerzy/BIB}

\newcommand{\here}{\vspace{1cm}}
\newcommand{\gone}[1]{}

\newcommand{\celegans}{{\it C. elegans} }
\newcommand{\CElegans}{\textit{Caenorhabditis elegans}}
\newcommand{\Celegans}{\textit{C.\ elegans}}

\newcommand{\CRep}{{\sc Cartesian-representation}}
\renewcommand{\CRep}{CR}
\newcommand{\CREP}{Cartesian-representation (CR)}

\newcommand{\sega}{{\textrm{1}}}
\newcommand{\segb}{{\textrm{2}}}
\newcommand{\segi}{i}
\newcommand{\external}{\textrm{ext}}
\newcommand{\actif}{{a}}
\newcommand{\actiF}{{A}}
\newcommand{\HR}{{R}}

\newcommand{\RB}{\textrm{rb}}
\newcommand{\HS}{\textrm{HSD}}
\newcommand{\minn}{\textrm{min}}
\newcommand{\maxx}{\textrm{max}}
\newcommand{\exact}{\textrm{CR}}

\newcommand{\curvature}{\kappa}
\newcommand{\curvatureWorm}{\curvature_{\textrm{w}}}

\newcommand{\arcLength}{s}
\newcommand{\arcLengthPrim}{\arcLength'}

\newcommand{\wormLength}{L}

\newcommand{\arcLengthSega}{\arcLength_\sega}
\newcommand{\arcLengthSegb}{\arcLength_\segb}
\newcommand{\arcLengthSegi}{\arcLength_\segi}
\newcommand{\arcLengthBeg}{\arcLength_0}

\newcommand{\lineS}{S}
\newcommand{\labCoordinateSystem}{{\cal O}}

\newcommand{\amplitude}{A}

\newcommand{\amplitudea}{A_1}
\newcommand{\amplitudeb}{A_2}
\newcommand{\amplitudei}{A_i}

\newcommand{\wavevector}{q}

\newcommand{\wavevectora}{q_1}
\newcommand{\wavevectorb}{q_2}
\newcommand{\wavevectori}{q_i}

\newcommand{\phase}{\phi}

\newcommand{\phasea}{\phi_1}
\newcommand{\phaseb}{\phi_2}
\newcommand{\phasei}{\phi_i}

\newcommand{\wavelength}{\lambda}
\newcommand{\wavelengthMax}{\lambda^\maxx}

\newcommand{\velocity}{v}
\newcommand{\snakyVelocity}{\velocity_s}

\newcommand{\forcei}[1]{{\bf F}_{#1}}
\newcommand{\torquei}[1]{{\bf T}_{#1}}

\newcommand{\dotproduct}{\boldsymbol{\cdot}}
\newcommand{\vecproduct}{\boldsymbol{\times}}

\newcommand{\bnabla}{\boldsymbol{\nabla}}

\newcommand{\longforce}{\zeta_{||}}
\newcommand{\transforce}{\zeta_{\perp}}
\newcommand{\lengthofchain}{N}
\newcommand{\wormlength}{L}
\newcommand{\forceratio}{\transforce/\longforce}

\newcommand{\Rcross}{\hat{\bf \Xi}}
\newcommand{\Rcrossi}{\Rcross_i}
\newcommand{\Rcrossj}{\Rcross_j}

\newcommand{\Nmin}{\lengthofchain_\minn}
\newcommand{\Nmax}{\lengthofchain_\maxx}

\newcommand{\longforceExact}[1]{\longforce^\exact(#1)}
\newcommand{\transforceExact}[1]{\transforce^{\exact}(#1)}
\newcommand{\longforceHSD}[1]{\longforce^\HS(#1)}
\newcommand{\transforceHSD}[1]{\transforce^\HS(#1)}

\newcommand{\totalForce}{{\bf F}}
\newcommand{\totalForceActive}{\totalForce^\actiF}
\newcommand{\totalForceHR}{\totalForce^\HR}

\newcommand{\totalTorque}{{\bf T}}
\newcommand{\totalTorqueActive}{\totalTorque^\actiF}
\newcommand{\totalTorqueHR}{\totalTorque^\HR}

\newcommand{\identity}{\hat{\bf I}}
\newcommand{\identityN}{\identity}

\newcommand{\zetaHSD}{\zeta_0}
\newcommand{\ZHStt}{\boldsymbol{\hat{\zeta}}^{\HS}}
\newcommand{\ZHStti}[1]{\ZHStt_{#1}}

\newcommand{\Z}{\boldsymbol{\zeta}}
\newcommand{\Zrs}{\boldsymbol{\zeta}^{r\actif}}
\newcommand{\Zts}{\boldsymbol{\zeta}^{t\actif}}
\newcommand{\Ztt}{\boldsymbol{\hat{\zeta}}^{tt}}
\newcommand{\Ztr}{\boldsymbol{\hat{\zeta}}^{tr}}
\newcommand{\Zrt}{\boldsymbol{\hat{\zeta}}^{rt}}
\newcommand{\Zrr}{\boldsymbol{\hat{\zeta}}^{rr}}
\newcommand{\Mtt}{\boldsymbol{\hat{\mu}}^{tt}}
\newcommand{\Mtr}{\boldsymbol{\hat{\mu}}^{tr}}
\newcommand{\Mrt}{\boldsymbol{\hat{\mu}}^{rt}}
\newcommand{\Mrr}{\boldsymbol{\hat{\mu}}^{rr}}

\newcommand{\velocityVec}{{\bf u}}
\newcommand{\velocityVeci}[1]{\velocityVec_{#1}}
\newcommand{\velocityVecSi}[1]{\velocityVec^\actiF_{#1}}
\newcommand{\barVelocityVecSi}[1]{\bar{\velocityVec}^\actiF_{#1}}
\newcommand{\velocityVecRB}{\velocityVec^\RB}
\newcommand{\velocityVecRBi}[1]{\velocityVec^\RB_{#1}}

\newcommand{\rotvelocity}{\omega}
\newcommand{\rotvelocityVec}{\boldsymbol{\rotvelocity}}
\newcommand{\rotvelocityVeci}[1]{\rotvelocityVec_{#1}}
\newcommand{\rotvelocitySi}[1]{\rotvelocity^\actiF_{#1}}
\newcommand{\rotvelocityVecSi}[1]{\rotvelocityVec^\actiF_{#1}}
\newcommand{\barRotvelocityVecSi}[1]{\bar{\rotvelocityVec}^\actiF_{#1}}
\newcommand{\rotvelocityVecRB}{\rotvelocityVec^\RB}
\newcommand{\rotvelocityVecRBi}[1]{\rotvelocityVec^\RB_{#1}}
\newcommand{\rodbetbeads}{\hat{\bf s}}
\newcommand{\rodbetbeadsi}[1]{\rodbetbeads_{#1}}
\newcommand{\rodrotvelocity}{\nu}
\newcommand{\rodrotvelocityi}[1]{\rodrotvelocity_{#1}}

\newcommand{\distance}{{\bf R}}
\newcommand{\distancei}[1]{\distance_{#1}}

\newcommand{\ex}{\hat{\bf e}_x}
\newcommand{\ey}{\hat{\bf e}_y}
\newcommand{\ez}{\hat{\bf e}_z}

\newcommand{\unittangentVector}{\hat{\bf t}}

\newcommand{\unittangentVectori}[1]{\unittangentVector_{#1}}

\newcommand{\unitnormalVector}{\hat{\bf n}}
\newcommand{\unitnormalVectori}[1]{\unitnormalVector_{#1}}

\newcommand{\externalPressure}{p^\external}
\newcommand{\externalPressurei}[1]{\externalPressure_{#1}}
\newcommand{\scatteredPressure}{p'}

\newcommand{\dipolarMoment}{{\bf D}}
\newcommand{\dipolarMomenti}[1]{\dipolarMoment_{#1}}

\newcommand{\lateralPosition}{\rho}
\newcommand{\lateralPositioni}[1]{\lateralPosition_{#1}}
\newcommand{\lateralPositionVector}{\boldsymbol{\lateralPosition}}
\newcommand{\lateralPositionVectori}[1]{\lateralPositionVector_{#1}}

\newcommand{\positiontensor}{\hat{\boldsymbol{g}}}

\newcommand{\snakyEfficiency}{\gamma_S}
\newcommand{\actualVelocity}{V_C}

\begin{document}

\title{Nematode Locomotion in Unconfined and Confined Fluids} 

\author{Alejandro Bilbao}
\affiliation{Department of Mechanical Engineering, Texas Tech University, Lubbock, Texas, United States of America}
\author{Eligiusz Wajnryb}
\affiliation{Institute of Fundamental Technological Research, Polish Academy of Sciences, Warsaw, Poland}
\author{Siva Vanapalli}
\affiliation{Department of Chemical Engineering, Texas Tech University, Lubbock, Texas, United States of America}
\author{Jerzy Blawzdziewicz}
\affiliation{Department of Mechanical Engineering, Texas Tech University, Lubbock, Texas, United States of America}

\date{\today}

\begin{abstract}
The millimeter-long soil-dwelling nematode \CElegans\ propels itself
by producing undulations that propagate along its body and turns by
assuming highly curved shapes. According to our recent study [PLoS ONE
  \textbf{7}, e40121 (2012)] all these postures can be accurately
described by a piecewise-harmonic-curvature (PHC) model.  We combine
this curvature-based description with highly accurate hydrodynamic
bead models to evaluate the normalized velocity and turning angles for
a worm swimming in an unconfined fluid and in a parallel-wall cell.
We find that the worm moves twice as fast and navigates more
effectively under a strong confinement, due to the large
transverse-to-longitudinal resistance-coefficient ratio resulting from
the wall-mediated far-field hydrodynamic coupling between body
segments.  We also note that the optimal swimming gait is similar to
the gait observed for nematodes swimming in high-viscosity fluids.
Our bead models allow us to determine the effects of confinement and
finite thickness of the body of the nematode on its locomotion. These
effects are not accounted for by the classical resistive-force and
slender-body theories.
\end{abstract}

\maketitle

\section{Introduction}

Locomotion of small swimming organisms
\cite{Lauga-Powers:2009,%
Cohen-Boyle:2010%
}
such as bacteria,\cite{Metcalfe-Pedley:2001} nematodes,
\cite{Juarez-Lu-Sznitman-Arratia:2010,%
Jung:2010,%
Sauvage-Argentina-Drappier-Senden-Simeon-DiMeglio:2011,%
Majmudar-Keaveny-Zhang-Shelley:2012} 
and planktonic species,\cite{Guasto-Rusconi-Stocker:2012} has
significant implications for diverse fields of study.  For example,
fundamental aspects of low-Reynolds-number locomotion are important
for understanding of long-range transport in swarms of swimmers,
\cite{Wu-Libchaber:2000,%
Hernandez-Ortiz-Stoltz-Graham:2005,%
  Underhill-Hernandez_Ortiz-Graham:2008,%
Koch-Subramanian:2011}
analysis of fouling of submerged surfaces due to formation of
bacterial biofilm, \cite{Pratt-Kolter:1998}
and description of evolutionary optimization.  \cite{Hosoi-Lauga:2010}
Locomotory mechanisms have also been harnessed in the design of
artificial swimmers
\cite{%
Zhang-Abbott-Dong-Kratochvil-Bell-Nelson:2009,%
Dreyfus-Baudry-Roper-Fermigier-Stone-Bibette:2005}
and functional microfluidic devices for biological assays.
\cite{Ahmed-Shimizu-Stocker:2010,%
Ma-Jiang-Shi-Qin-Lin:2009,%
Chung-Crane-Lu:2008,Chronis-Zimmer-Bargmann:2007,%
Lockery-Lawton-Doll-Faumont-Coulthard-Thiele-Chronis-McCormick-Goodman-Pruitt:2008} 

A number of recent locomotion studies focused on a submillimeter-size
nematode \CElegans.
\cite{Juarez-Lu-Sznitman-Arratia:2010,%
Jung:2010,%
Sznitman-Shen-Purohit-Arratia:2010a,%
Sznitman-Shen-Sznitman-Arratia:2010,%
Sauvage-Argentina-Drappier-Senden-Simeon-DiMeglio:2011,%
Majmudar-Keaveny-Zhang-Shelley:2012,%
Korta-Clark-Gabel-Mahadevan-Samuel:1007,%
Berri-Boyle-Tassieri-Hope-Cohen:2009,%
Fang_Yen-Wyart-Xie-Kawai-Kodger-Chen-Wen-Samuel:2010,%
Berman-Kenneth-Sznitman-Leshansky:2013%
}
This soil-dwelling worm is a model organism for investigations of
genetic regulation and neural control of muscular activity,
\cite{Etheridge-Oczypok-Lehmann-Fields-Shephard-Jacobson-Szewczyk:2012}
motion,
\cite{Omura-Clark-Samuel-Horvitz:2012,%
Ward-Walker-Feng-Xu:2009} 
and behavior.
\cite{Downes-Birsoy-Chipman-Rothman:2012,
McCormick-Gaertner-Sottile-Phillips-Lockery:2011,%
Gray-Hill-Bargmann:2005}
Quantitative understanding of nematode locomotion is thus important
for the mutant testing and analysis of neuro--muscular system.  Drug
screening assays also benefit from locomotion investigations since
motility of \Celegans\ is often used as a phenotypic readout for drug
efficacy.  \cite{Artal_Sanz-deJong-Tavernarakis:2006}

\begin{figure}[b]
\includegraphics{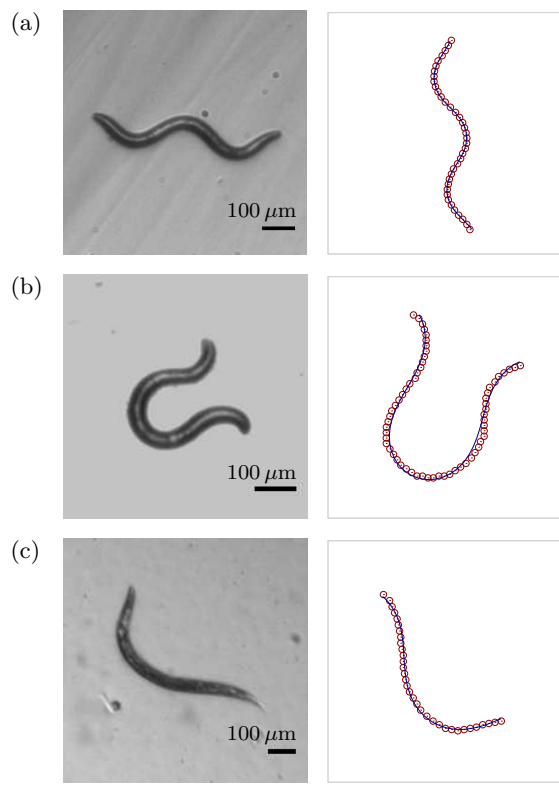}
\caption{Typical body postures of \celegans (a) crawling on agar
  ($W$-shape), (b) making $\Omega$-turn on agar, and (c) swimming in
  water ($C$-shape).  The corresponding right panels show PHC
  description of the shapes in the left panels.  Circles represent the
  numerical skeletons of the worm images, and the lines show results of
  a single-mode PHC model (see Sec.\ \ref{kinematics} and Ref.\
\onlinecite{%
Padmanabhan-Khan-Solomon-Armstrong-Rumbaugh-Vanapalli-Blawzdziewicz:2012}).
}
\label{real worms and fits}
\end{figure}

\Celegans\ propels itself by producing sinuous undulation propagating
along the body, \cite{Gray-Lissmann:1964} and turns by assuming
strongly curved $\Omega$- and loop-shaped body
postures.\cite{Gray-Hill-Bargmann:2005} The wavelength of the
undulation depends on the environment in which the nematode moves.
Crawling on smooth surfaces (such as agar in laboratory experiments),
\Celegans\ assumes short-wave $W$-shaped body postures, and during
swimming in water it displays a longer-wave $C$-shaped body form.
Using a simple set of body movements, nematodes efficiently maneuver
in diverse environments such as soft moist surfaces, bulk and confined
fluids, and complex inhomogeneous media.  Typical body shapes of
\Celegans\ are illustrated in Fig.~\ref{real worms and fits}.

% FIGURE 1

Recent  investigations of the response of \Celegans\ to increased
fluid viscosity
\cite{Korta-Clark-Gabel-Mahadevan-Samuel:1007,%
Berri-Boyle-Tassieri-Hope-Cohen:2009,%
Fang_Yen-Wyart-Xie-Kawai-Kodger-Chen-Wen-Samuel:2010,%
Vidal_Gadea-Topper-Young-Crisp-Kressin-Elbel-Maples-Brauner-Erbguth-Axelrod-Gottschalk-Siegel-Pierce_Shimomura:2011}
and confinement pressure 
\cite{Lebois-Sauvage-Py-Cardoso-Ladoux-Hersen-Di_Meglio:2012}
have shown a continuous transition between the long-wave swimming gait
in water and a short-wave gait (similar to the crawling gait) in
high-viscosity fluids.  Understanding of this phenomenon will provide
important clues for modeling neural control and biomechanics. 

To elucidate this transition, a detailed analysis of hydrodynamics of
swimming for a variety of gaits is needed.  Using numerical models,
this study investigates nematode swimming in different environments.
Since experiments are often performed in parallel-wall cells,
\cite{Jung:2010,%
Sznitman-Shen-Sznitman-Arratia:2010,%
Lebois-Sauvage-Py-Cardoso-Ladoux-Hersen-Di_Meglio:2012}
we consider swimming in bulk fluids and in fluids confined between two
parallel walls.  The analysis draws on our recently developed
piecewise-harmonic-curvature (PHC) description of worm kinematics.
\cite{Padmanabhan-Khan-Solomon-Armstrong-Rumbaugh-Vanapalli-Blawzdziewicz:2012}
As discussed in Sec.\ \ref{kinematics}, the PHC model allows us to
quantitatively describe nematode shapes used in crawling and
swimming. It is also a convenient tool to investigate turns.

In Sec.\ \ref{Nematode hydrodynamics} we formulate a mobility relation
for active-particle locomotion, present our hydrodynamic modeling
techniques, and discuss the role of confinement in generating
hydrodynamic propulsive force.  The results for swimming velocity for
different nematode gaits are given in Sec.\ \ref{efficiency}.  Our
analysis of turning maneuvers presented in Sec.\ \ref{Hydrodynamics of
  Turns} (to our knowledge the first quantitative study of the
hydrodynamics of nematode turns) will have important implications for
investigations of nematode chemotaxis.  \cite{Albrecht-Bargmann:2011}

\section{Nematode kinematics: piecewise-harmonic-curvature representation 
of nematode gait}
\label{kinematics}

% FIGURE 2
\begin{figure}
\includegraphics{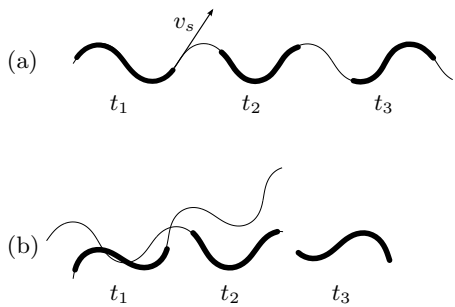}
\caption{Time progressions for a crawling and a swimming nematode
  performing the same set of body movements: (a) crawling worm (thick
  line) slides with velocity $\snakyVelocity$ along a predetermined
  curve (thin line); (b) swimming worm undergoes translational and
  rotational slip superposed with the motion along the curve.}
\label{snapshot sequence crawling versus swimming}
\end{figure}

% FIGURE 3
\begin{figure}[b]
\includegraphics{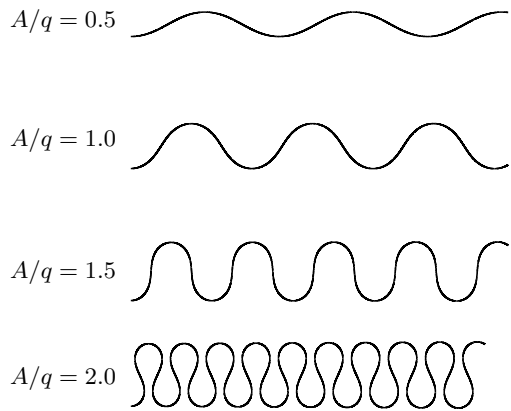}
\caption{Curves defined by sinusoidal curvature \eqref{harmonic
    curvature single} for several values of the normalized amplitude
  $\amplitude/\wavevector$.}
\label{various curve shapes}
\end{figure}

\Celegans\ move forward by producing sinuous undulations and turn by
assuming strongly curved body shapes.  The deformations of the
nematode body take place in two dimensions, i.e., in the
ventral--dorsal plane.  The head of the nematode can additionally move
normal to this plane, resulting in three-dimensional motion.  Here we
restrict our analysis to two-dimensional swimming; the effects of
normal movements of nematode head will be described elsewhere.

 We have recently demonstrated that the gait of crawling and
swimming \Celegans\ can be accurately modeled using a
piecewise-harmonic expression for its body curvature.
\cite{%
Padmanabhan-Khan-Solomon-Armstrong-Rumbaugh-Vanapalli-Blawzdziewicz:2012}
This PHC representation applies to individual body shapes (including
$W$-shapes, $\Omega$-shapes, and $C$-shapes depicted in
Fig.\ \ref{real worms and fits}) as well as to whole tracks of
\Celegans\ crawling on agar.  We have also shown that a similar
description applies to swimming \Celegans.

According to the PHC model, the evolution of the body shape is
described by the 
curvature wave propagating along the
nematode body,
\begin{equation}
\label{curvature wave}
\curvatureWorm(\arcLengthPrim,t)=\curvature(\arcLengthPrim+\snakyVelocity t),
\end{equation}
where $\snakyVelocity$ is the wave-propagation velocity, $t$ is time,
and $\arcLengthPrim$ is the coordinate along the nematode body (with
$\arcLengthPrim=0$ and $\arcLengthPrim=L$ denoting the tail and head
positions).  The analysis of nematode body postures 
\cite{%
Padmanabhan-Khan-Solomon-Armstrong-Rumbaugh-Vanapalli-Blawzdziewicz:2012}
shows that the
curvature wave is well represented by the piecewise-harmonic function
\begin{equation}
\curvature(\arcLength)=\left\{\begin{matrix}
   \amplitudea\cos(\wavevectora\arcLength+\phasea),
         \qquad \arcLengthBeg \le \arcLength \le \arcLengthSega, \\
   \amplitudeb\cos(\wavevectorb\arcLength+\phaseb),
         \qquad \arcLengthSega \le \arcLength \le \arcLengthSegb, \\
\hspace{\fill}\ldots\hspace{\fill}
\hphantom{\qquad \arcLengthSega \le \arcLength \le \arcLengthSegb} 
\end{matrix}\right. \\
\label{harmonic curvature multiple}
\end{equation}
with
\begin{equation}
\arcLength=\arcLengthPrim+\snakyVelocity t,
\label{definition of s in relation to time}
\end{equation}
where
$\arcLengthSegi$ are the mode-change points, and $\amplitudei$,
$\wavevectori$, and $\phasei$ are the amplitude, wavevector, and phase
of the mode $i=1,2,\ldots\,$  

The real-space curves defined by the
curvature $\curvature(\arcLength)$ can be obtained by solving
Frenet--Serret equations
\begin{equation}
\label{curvature diffyQ}
  \left[ 
    \begin{array}{cc} 
      \ddot{x}(\arcLength)\\ 
      \ddot{y}(\arcLength) 
    \end{array} 
  \right] =
     \curvature(\arcLength) 
  \left[ 
    \begin{array}{cc} 
      -\dot{y}(\arcLength)\\
       \dot{x}(\arcLength)
    \end{array}
  \right],
\end{equation}
where $x$ and $y$ are the Cartesian coordinates in the plane of
motion, and the dot denotes differentiation with respect to the
variable $s$.  

According to equation \eqref{curvature wave}, a sequence of nematode's
body postures corresponds to a line section of length $L$ sliding with
velocity $\snakyVelocity$ along a curve defined by
Eq.\ \eqref{curvature diffyQ}.  In addition to sliding along the
curve, a swimming nematode undergoes a rigid-body translational and
rotational slip with respect to the surrounding medium.  This dynamics
is schematically illustrated in Fig.\ \ref{snapshot sequence crawling
  versus swimming}.

The family of curves defined by the single-mode relation
\begin{equation}
\label{harmonic curvature single}
\kappa(s)=\amplitude\cos(\wavevector\arcLength+\phase)
\end{equation}
with different values of the normalized amplitude
$\amplitude/\wavevector$ is illustrated in Fig.\ \ref{various curve
  shapes}.  Changing $\wavevector$ at a fixed value of
$\amplitude/\wavevector$ results in rescaling of the whole curve.
Single-mode fits \eqref{harmonic curvature single} to the nematode
shapes depicted in the left panels of Fig.~\ref{real worms and fits}
are presented in the corresponding right panels.  We note that the
shape of a nematode of length $\wormLength$ at time $t$ is described
by three dimensionless parameters: the normalized amplitude
$\amplitude/\wavevector$, dimensionless wavevector
$\wavevector\wormLength$, and the phase
$\phi'=\phi+\wavevector\snakyVelocity t$.

As argued in Ref.\
\onlinecite{%
Padmanabhan-Khan-Solomon-Armstrong-Rumbaugh-Vanapalli-Blawzdziewicz:2012},
nematodes use a single mode to move forward and they switch modes in
order to turn.  Our analysis of rectilinear swimming (see
Sec.\ \ref{efficiency}) explores the entire space of the PHC
parameters $\amplitude/\wavevector$ and $\wavevector\wormLength$.  We
determine the dependence of the normalized swimming velocity on the
nematode gait, thus providing important insights regarding the gait
transition observed for \celegans swimming in highly viscous fluids.
\cite{Korta-Clark-Gabel-Mahadevan-Samuel:1007,%
Berri-Boyle-Tassieri-Hope-Cohen:2009,%
Fang_Yen-Wyart-Xie-Kawai-Kodger-Chen-Wen-Samuel:2010,%
Vidal_Gadea-Topper-Young-Crisp-Kressin-Elbel-Maples-Brauner-Erbguth-Axelrod-Gottschalk-Siegel-Pierce_Shimomura:2011}
In our discussion of turning maneuvers (see Sec.\ \ref{Hydrodynamics
  of Turns}) we use a more limited set of parameters, because in a
multi-mode system the parameter space is too large to be fully
explored.  Assuming that the nematode switches from the default
$W$-shaped or $C$-shaped forward-locomotion mode [cf.\ Figs.\ \ref{real
    worms and fits}(a) and  \ref{real worms and fits}(c)] to
$\Omega$-shaped turning mode [cf.\ Fig.\ \ref{real worms and fits}(b)]
and then reverts to the default mode, we focus on the dependence of
the turning angle on the phases at which the PHC modes are switched.

\section{Nematode hydrodynamics}
\label{Nematode hydrodynamics}

\subsection{Balance of forces and torques acting on the nematode body}
\label{Force Balance}

 In our model, the undulating body of the nematode experiences
hydrodynamic forces and torques produced by the propagating wave of
PHC. Under creeping flow conditions (assumed herein), the total
hydrodynamic force $\totalForce$ and torque $\totalTorque$ acting on a
swimming nematode can be expressed as a superposition of the
contribution produced by the predetermined motion with velocity
$\snakyVelocity$ along the line defined by the PHC relation
\eqref{harmonic curvature multiple} and Frenet--Serret equations
\eqref{curvature diffyQ} [cf.\ Fig.\ \ref{snapshot sequence crawling
    versus swimming}(a)] and the contribution due to the rigid-body
translation and rotation with the linear and angular velocities
$\velocityVecRB$ and $\rotvelocityVecRB$ [cf.\ Fig.\ \ref{snapshot
    sequence crawling versus swimming}(b)].  In the creeping flow
regime both these terms are given by the linear friction relations.
The total force and torque balance can thus be expressed as
\begin{equation}
\label{total force of a body}
\begin{bmatrix} 
  \totalForce \\ 
  \totalTorque 
\end{bmatrix}
  =
\begin{bmatrix} 
  \Zts \\ 
  \Zrs  
\end{bmatrix}
    \snakyVelocity
  +
\begin{bmatrix}
  \Ztt & \Ztr \\
  \Zrt & \Zrr
\end{bmatrix}
  \cdot
\begin{bmatrix} 
  \velocityVecRB \\
  \rotvelocityVecRB 
\end{bmatrix},
\end{equation} 
where $\Z^{\alpha\actif}$ ($\alpha=t,r$) are the active-force- and
active-torque-generation tensors (with superscript $a$ referring to
the active contribution), and $\Z^{\alpha\beta}$ ($\alpha,\beta=t,r$)
are the translational and rotational hydrodynamic resistance tensors.
All the above tensors depend on the instantaneous posture of the
nematode body.  Since the nematode is force- and torque-free,
	\begin{equation}
\begin{bmatrix} 
  \totalForce \\ 
  \totalTorque 
\end{bmatrix}
=0,
\label{total force of a body to zero}
\end{equation}
the friction relation \eqref{total force of a body} yields the
following mobility relation for the rigid-body translation and
rotation of an active particle
\begin{equation}
\begin{bmatrix} 
  \velocityVecRB \\ 
  \rotvelocityVecRB 
\end{bmatrix}
  =-
\begin{bmatrix}
  \Mtt & \Mtr \\
  \Mrt & \Mrr
\end{bmatrix}
  \cdot
\begin{bmatrix} 
  \Zts \\ \Zrs 
\end{bmatrix} 
\snakyVelocity,
\label{rigid body velocity}
\end{equation}
where
\begin{equation}
\begin{bmatrix}
  \Mtt & \Mtr \\
  \Mrt & \Mrr
\end{bmatrix}
  =
\begin{bmatrix}
  \Ztt & \Ztr \\
  \Zrt & \Zrr
\end{bmatrix}^{-1}
\label{mobility matrix definition}
\end{equation}
is the mobility matrix for a given nematode posture.  In
Secs.\ \ref{Bead Models} and \ref{Hydrodynamic Induced Transversal and
  Longitudinal Forces} we describe our methods for evaluating the
above matrices using accurate bead-chain models.

% FIGURE 4
%\input{figtex/beadmodelworm.tex} %label= bead model
\begin{figure}
	\includegraphics{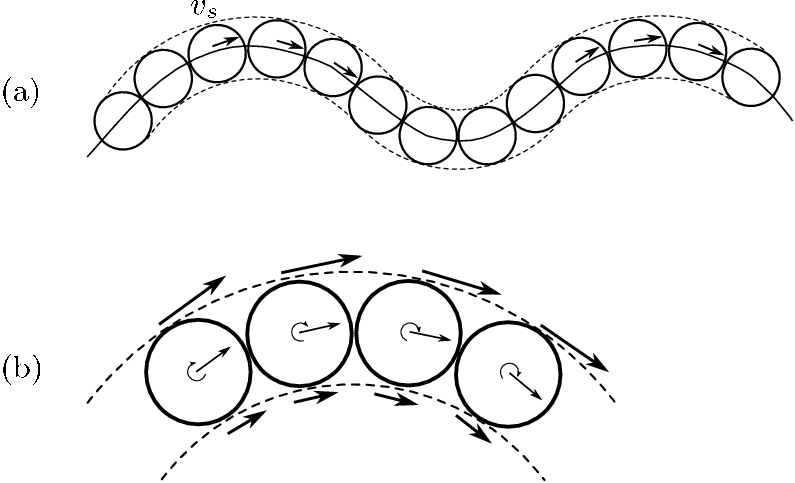}
\caption{ Nematode body modeled as a chain of touching
spheres. (a) The spheres follow the curve defined by the PHC model
with wave-velocity $\snakyVelocity$. (b) Prescribed individual bead
rotations mimic the motion of the interface of the nematode. }
\label{bead model}
\end{figure}

\subsection{Bead models}
\label{Bead Models}

 To determine the mobility matrix \eqref{mobility matrix
  definition} we model the nematode as an active chain of
hydrodynamically interacting spheres.  The chain performs a set of
motions similar to a sequence of body postures of a real nematode [see
  Fig.\ \ref{bead model}(a)].  In addition to the translational
motion, the beads rotate to mimic deformation of the interface of an
elongated body, as illustrated in Fig. \ref{bead model}(b).  In more
detail, the bead-chain kinematics is described in
Appendix~\ref{Bead-chain model}. 

\here

For each chain configuration (in our simulations described by the PHC
model), the active-force and friction tensors $\Z^{\alpha\beta}$ in
Eq.\ \eqref{total force of a body} are evaluated from the
corresponding hydrodynamic-resistance matrix for a system of
hydrodynamically coupled spheres.  The bead positions are then updated
according to the changing chain configuration and the rigid-body
velocity \eqref{rigid body velocity}.  In our simulations we use the
forth-order Runge--Kutta method for time stepping.

It has been shown that bead-chain
models faithfully reproduce hydrodynamic interactions of elongated
bodies,
\cite{Guzowski-Cichocki-Wajnryb-Abade:2008}
so we expect that our calculations accurately describe nematode motion.
Details of the chain kinematics and the relevant resistance relations
are presented in Appendix \ref{Bead-chain model}.

We consider a nematode swimming in two distinct geometries:
(\textit{a}) an unbounded fluid and (\textit{b}) the midplane of a
parallel-wall channel.  In the first case, the hydrodynamic
interactions between the beads representing the nematode body are
evaluated using the {\sc Hydromultipole} method.
\cite{Cichocki-Felderhof-Hinsen-Wajnryb-Blawzdziewicz:1994} In the
second case we apply the {\CREP} method,
\cite{Bhattacharya-Blawzdziewicz-Wajnryb:2005a,%
Bhattacharya-Blawzdziewicz-Wajnryb:2005}
and we also employ a computationally efficient approximate
Hele--Shaw-dipole (HSD) method (see Appendix \ref{Hele--Shaw point
  dipole approximation}).  

 In our simulations of swimming nematodes we use chains of length
$N=30$ beads, consistent with the average thickness-to-length ratio of
\Celegans.  Bead models allow us to determine effects of finite body
thickness and confining walls on the nematode locomotion.  These
effects are not accounted for in the standard resistive force theory
(RFT)
\cite{Gray-Hancock:1955,Johnson-Brokaw:1979} 
and slender-body theory (SBT),
\cite{Johnson:1980,Johnson-Brokaw:1979} 
and a modified SBT for a confined cylinder between parallel
walls\cite{Katz-Blake-Paveri-Fontana:1975} is inaccurate for
geometries relevant for nematode locomotion, as discussed in
Sec.\ \ref{Hydrodynamic Induced Transversal and Longitudinal Forces}.

% FIGURE 5
\begin{figure}
\includegraphics{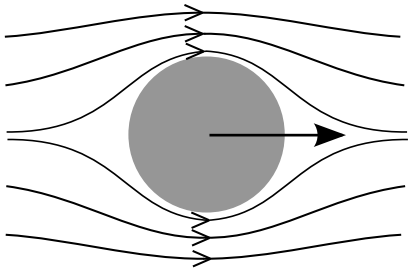}
\caption{Schematic of the flow field generated by an elongated body
  dragged in a transverse direction through an unconfined
  fluid. Overall, the scattered flow is in the same direction as the
  velocity of the body; the resulting resistance force is moderate.}
	\label{unConfined cylinder flow field}
\end{figure}

% FIGURE 6
\begin{figure}
\includegraphics{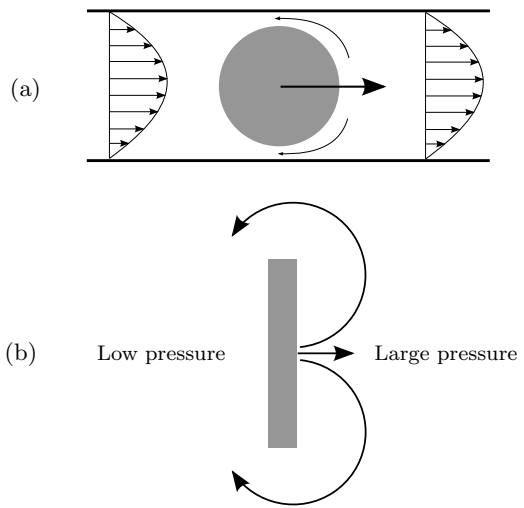}
\caption{The elongated-piston effect: The motion of an
  elongated body dragged in the transverse direction in a
  parallel-wall channel produces long-range pressure-driven
  recirculation pattern.  The corresponding pressure drop across the
  body results in a large resistance force.  (a) Side view of the
  system and (b) top view.}
\label{Confined cylinder flow field}
\end{figure}

% FIGURE 7
\begin{figure}	
\includegraphics{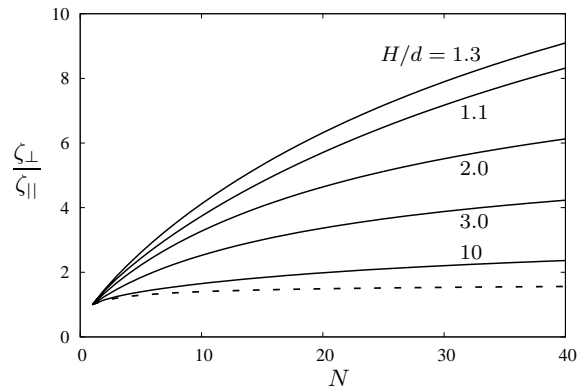}
\caption{Ratio $\transforce/\longforce$ between the transverse and
  longitudinal hydrodynamic-resistance coefficients for a linear chain
  of equal-size spheres vs.\ the chain length $\lengthofchain$ for
  unconfined system (dotted line) and parallel-wall channels (solid
  lines).  Channel width normalized by the bead diameter $H/d$, as
  labeled.  The chain moves in the midplane of the channel.}
\label{force ratio of confined and free swimming}
\end{figure}

% FIGURE 8
% \input{figtex/FratVSHod.tex} % label= Force ratio versus gap width
\begin{figure}
\includegraphics{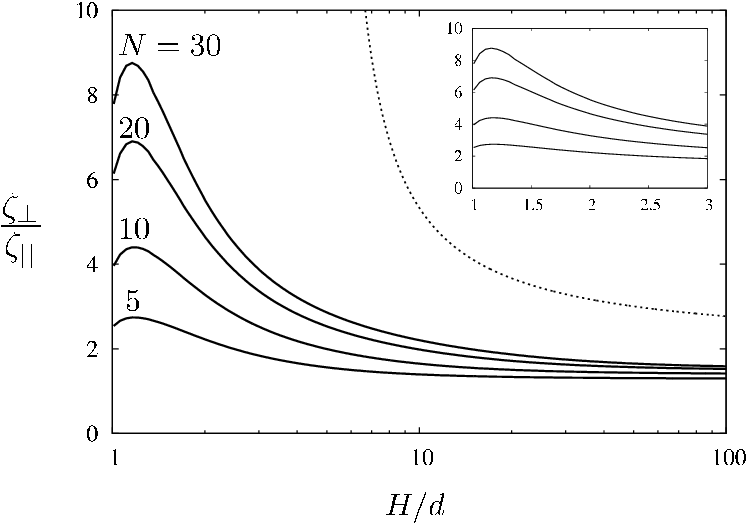}
\caption{ Ratio $\forceratio$ between transverse and logitudinal
  hydrodynamic-resistance coefficients vs. normalized gap width $H/d$.
  Linear chain of equal-size spheres (solid lines) of chain lengths as
  labeled.  Results of a modified slender-body theory
  \cite{Katz-Blake-Paveri-Fontana:1975} for a confined cylinder of
  diameter $d$ and infinite length (dotted line). Inset shows a blowup
  of the region of the moderate values of $H/d$. }
\label{Force ratio versus gap width}
\end{figure}

\subsection{Effect of confinement on transverse and longitudinal hydrodynamic forces}
\label{Hydrodynamic Induced Transversal and Longitudinal Forces}

Effective undulatory locomotion requires large ratios
$\transforce/\longforce$ between the transverse and longitudinal
resistance coefficients $\transforce$ and $\longforce$ that describe
friction forces acting on segments of an elongated body.
\cite{Gray-Hancock:1955,%
Sauvage-Argentina-Drappier-Senden-Simeon-DiMeglio:2011} 
 As demonstrated in our earlier studies,
\cite{Bhattacharya-Blawzdziewicz-Wajnryb:2005a,%
Bhattacharya-Blawzdziewicz-Wajnryb:2005} 
(also see Refs.\ \onlinecite{Katz-Blake-Paveri-Fontana:1975} and
\onlinecite{Han-Alsayed-Nobili-Zhang-Lubensky-Yodh-2006}), the
resistance-coefficient ratio is significantly affected by confinement,
and can be several times larger in a parallel-wall channel than in an
unconfined fluid.  Here we show that confinement enhances the
efficiency of undulatory swimming as a result of this increased
resistance-coefficient ratio.

To elucidate the effect of confinement on the propulsion forces in
undulatory swimming, we consider the flow field produced by a rigid
elongated body in transverse motion through an unconfined fluid and
along a flat parallel-wall channel.  As schematically illustrated
in Fig.\ \ref{unConfined cylinder flow field}, an unconfined cylinder
produces a long-range flow in the same overall direction as the
velocity of the cylinder.  The resulting transverse resistance force
is relatively low.  In the limit of infinite cylinder length, the
transverse-to-longitudinal resistance-coefficient ratio
logarithmically approaches the asymptotic value $\forceratio=2$,
\cite{Weinberger:1972,%
Blawzdziewicz-Wajnryb-Given-Hubbard:2005}
and for finite cylinders it is even smaller.

In contrast, for a long cylinder strongly confined between two
parallel walls (as shown in Fig.\ \ref{Confined cylinder flow field})
the transverse resistance is significantly larger.  This large
resistance stems from the pressure distribution needed to drive fluid
flow from the region in front of the cylinder to the region behind it.
\cite{Bhattacharya-Blawzdziewicz-Wajnryb:2005a,%
Bhattacharya-Blawzdziewicz-Wajnryb:2005,%
Han-Alsayed-Nobili-Zhang-Lubensky-Yodh-2006}
For a tightly confined system, the fluid is forced around the ends of
the cylinder, as opposed to leaking between the cylinder and the walls
[cf.\ Fig.\ \ref{Confined cylinder flow field}(b)].  Thus the cylinder
acts as a piston pushing fluid in the direction parallel to the walls.
The parabolic flow produced by this elongated-piston effect occurs
over a distance comparable to the cylinder length and thus requires a
large pressure drop, producing a large resistance force.

The effect of the above hydrodynamic-friction mechanism on the
resistance-coefficient ratio $\transforce/\longforce$ is depicted in
Figs.\ \ref{force ratio of confined and free swimming} and \ref{Force
  ratio versus gap width} where we show $\transforce/\longforce$ for
submerged linear chains of spheres moving in the midplane of a
parallel-wall channel.  In Fig.\ \ref{force ratio of confined and free
  swimming} the resistance-coefficient ratio is plotted vs. the chain
length $\lengthofchain$ for different confinement ratios $H/d$ (where
$H$ is the channel width and $d$ is the bead diameter).  Figure
\ref{Force ratio versus gap width} depicts the dependence of
$\transforce/\longforce$ on $H/d$ for varying chain lengths. In
addition, Fig.\ \ref{Force ratio versus gap width} also compares our
bead-chain results with the resistance-coefficient ratio calculated
using a modified SBT for a confined cylinder with $d\ll H\ll l$, where
$d$ is the cylinder diameter and $l$ is the cylinder length.
\cite{Katz-Blake-Paveri-Fontana:1975} We find that SBT significantly
overpredicts the resistance ratio for the systems considered in our
study, most likely due to large logarithmic corrections resulting in a
narrow validity range of the approximation.

Our calculations show that, consistent with the elongated-piston
mechanism, the resistance-coefficient ratio increases with the
increasing chain length.  The largest ratio is observed at a
dimensionless channel separation of $H/d\approx1.3$. For tighter
confinements the resistance forces are dominated by the lubrication
forces between the walls and the particle.  The lubrication forces are
significantly more isotropic
\cite{Sauvage-Argentina-Drappier-Senden-Simeon-DiMeglio:2011} than the
forces associated with the piston effect; hence the decrease of the
friction-coefficient ratio for $H/d<1.3$.  For larger wall
separations, more fluid leaks between the particle and walls, which
also results in a decrease of $\transforce/\longforce$.  The decrease
is gradual: we find that the elongated-piston effect is fairly strong
for $H/d=3$, and the enhanced transverse resistance is still
noticeable even for $H/d=10$.

The above results suggest that the effectiveness of nematode swimming
is strongly affected by confinement.  This conclusion is supported by
explicit calculations presented in Sec.\ \ref{efficiency}.

\section{Effectiveness of locomotion for different confinements and 
nematode gaits}
\label{efficiency}

% FIGURE 9
\begin{figure}
\includegraphics{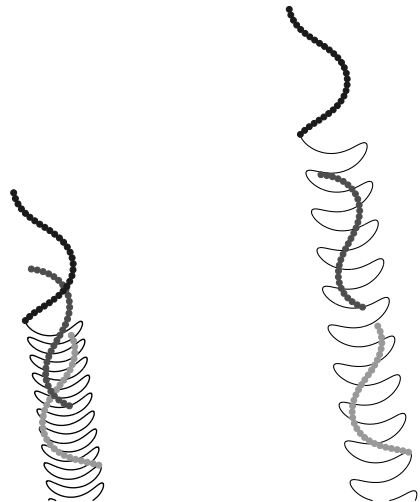}
\caption{Tail trajectories and snapshots of body positions at equally
  spaced times for a nematode swimming in unconfined fluid (left) and
  in the midplane of a parallel-wall channel of normalized width
  $H/d=1.3$ (right). }
\label{swimming worm trajectories}
\end{figure}

% FIGURE 10
%\input{figtex/GammaVSq_varA_bulk.tex}
\begin{figure}
\includegraphics{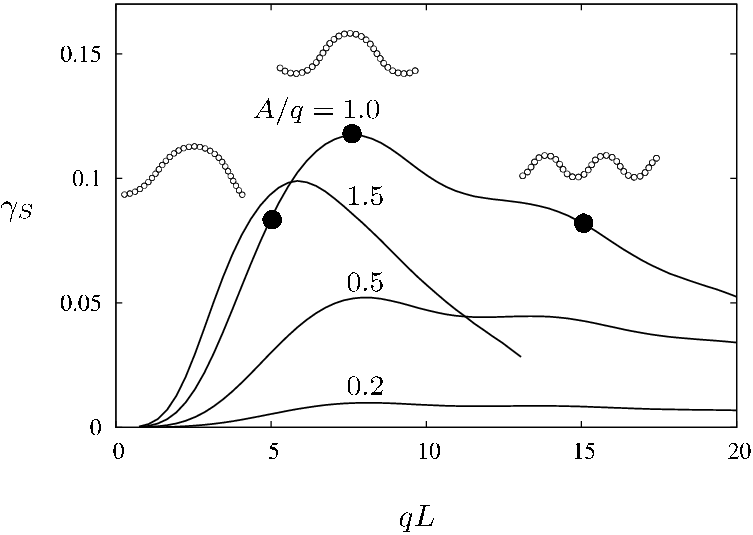}
\caption{Normalized swimming velocity $\snakyEfficiency$ vs. wave
vector $\wavevector$ normalized by the worm length $\wormlength$ for
a nematode swimming in an unconfined fluid.  Normalized amplitude
$\amplitude/\wavevector$ as labeled.  Insets show nematode shapes
for parameters corresponding to the points indicated by filled
circles.}
\label{gammaS vs q in the bulk}
\end{figure}

% FIGURE 11
% \input{figtex/GSvQL_helevCR.tex} %label= panel: gammaS vs q
\begin{figure}
	\includegraphics{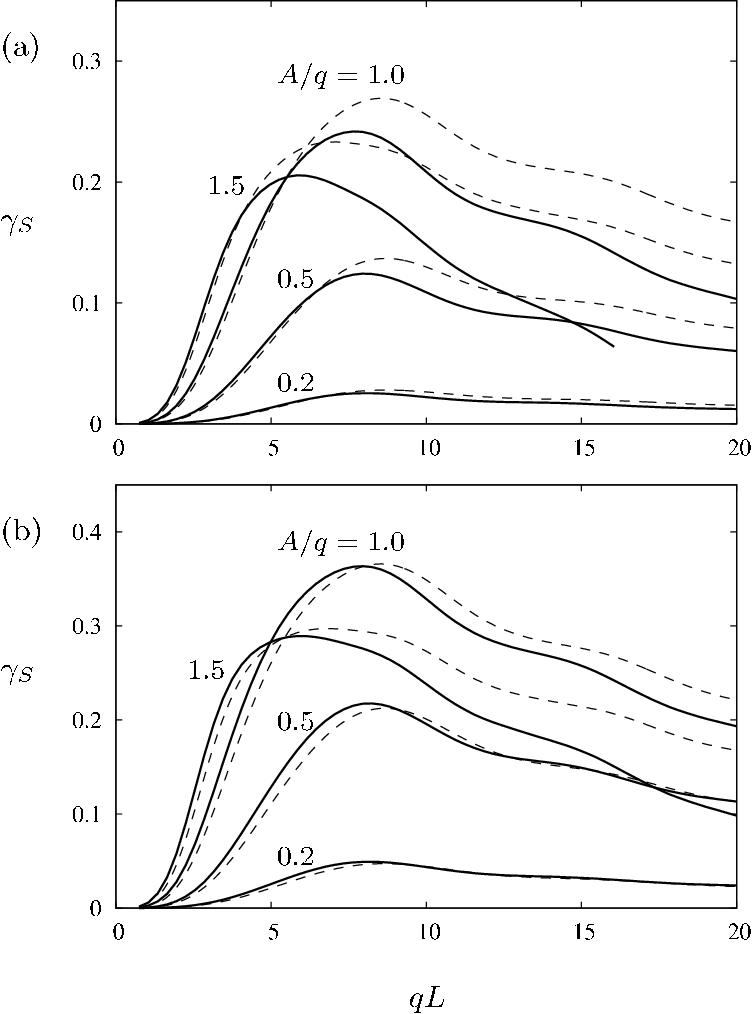}
\caption{ Normalized swimming velocity $\snakyEfficiency$ vs. wave
vector $\wavevector$ normalized by the worm length $\wormlength$
for a nematode swimming in a midplane of a parallel-wall channel of
width (a) $H/d=3$ and (b) $H/d=1.3$.  Normalized amplitude
$\amplitude/\wavevector$ as labeled.  Results obtained using
{\CRep} method (solid lines) and HSD approximation (dashed lines).
}
\label{panel: gammaS vs q}
\end{figure}

% FIGURE 12
% \input{figtex/GammaVSAq_varH_all.tex}
\begin{figure}
	\includegraphics{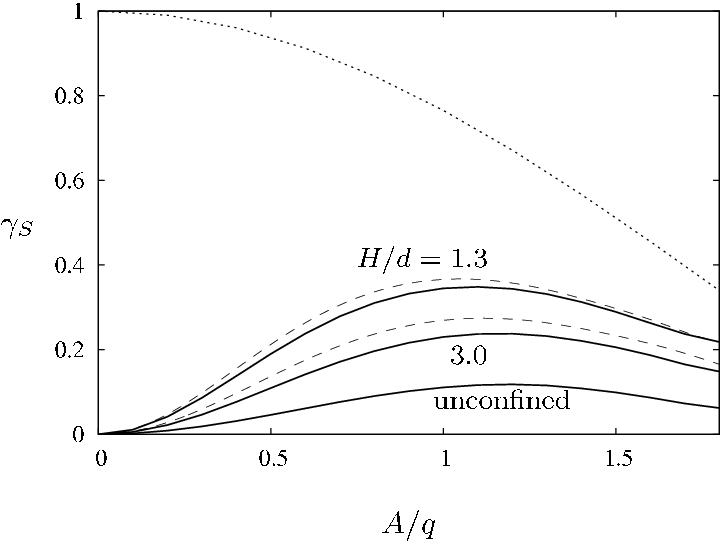}
\caption{ Normalized velocity $\snakyEfficiency$ vs. normalized
 amplitude $A/q$ for a nematode swimming in unconfined fluid
 (as labeled) and in the midplane of a parallel-wall channel
 for channel width as labeled.  Results obtained using
 {\CRep} method (solid lines) and HSD approximation (dashed
 lines). The normalized wavevector $\wavevector\wormlength$
 corresponds to the maximal efficiency for the given geometry
 and amplitude.  Dotted line represents normalized velocity
 for a nematode crawling without slip. }
\label{gammaS vs Aq all}
\end{figure}

% FIGURE 13
% \input{figtex/GammaVSH_CR_hele.tex}
\begin{figure}
	\includegraphics{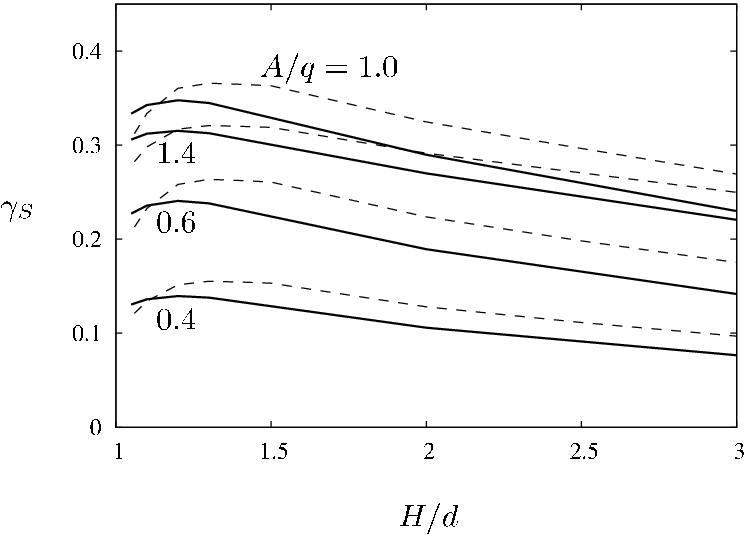}
\caption{ Normalized velocity $\snakyEfficiency$ vs.\ normalized
channel width $H/d$ for a nematode swimming in the midplane of a
parallel-wall channel for normalized amplitudes
$\amplitude/\wavevector$ as labeled.  Results obtained using {\CRep}
method (solid lines) and HSD approximation (dashed lines). The
normalized wavevector $\wavevector\wormlength$ corresponds to the
maximal efficiency for a given channel geometry and normalized
amplitude. }
\label{GammaS vs H varA}
\end{figure}

Figure \ref{swimming worm trajectories} illustrates typical
trajectories of a swimming nematode in unconfined (left) and confined
(right) fluid.  In both cases the nematode uses the same sequence of
body motions.  Each panel shows the path of the nematode tail and
three snapshots of body postures separated by the same phase
difference.  

Consistent with the discussion in Sec.\ \ref{Hydrodynamic
  Induced Transversal and Longitudinal Forces}, the unconfined
nematode experiences much more backward slip with respect to the
surrounding fluid (compared to the motion with no sidewise slip along
the harmonic-curvature path).   Hence, the worm is less efficient,
i.e., it moves a shorter distance for the same sequence of body
postures than the confined nematode.  Our numerical results closely
resemble trajectories of swimming \celegans depicted in Fig.\ 1(a) of
Ref.\ \onlinecite{Sznitman-Shen-Sznitman-Arratia:2010}.

 In order to quantify the effect of the hydrodynamic slip on the
nematode motion, we define the normalized swimming velocity
\begin{equation}
  \snakyEfficiency = \frac{\actualVelocity}{\snakyVelocity},
\label{kinematic efficiency of swimming worm}
\end{equation}
where $\actualVelocity$ denotes the average velocity of the worm, and
$\snakyVelocity$ is the propagation velocity of the curvature wave
\eqref{curvature wave} along the nematode body. 

Our results for the normalized velocity \eqref{kinematic efficiency of
  swimming worm} for different swimming gaits of confined and
unconfined nematodes are presented in Figs.\ \ref{gammaS vs q in the
  bulk}--\ref{GammaS vs H varA}.  The nematode gait is characterized
by the dimensionless amplitude $\amplitude/\wavevector$ (which defines
the shapes of the no-slip trajectories represented in
Fig.\ \ref{various curve shapes}), and by the wavevector $\wavevector$
normalized by the nematode length $\wormlength$.

Figures \ref{gammaS vs q in the bulk} and \ref{panel: gammaS vs q}
show the swimming velocity $\snakyEfficiency$ vs.\ the normalized
wavevector $\wavevector\wormlength$ in unconfined and confined
systems, respectively, for several values of the dimensionless
amplitude.   The insets in Fig.\ \ref{gammaS vs q in the bulk}
show the nematode shapes corresponding to characteristic parameter
values.  Since the parameter ranges in Figs. \ref{gammaS vs q in the
  bulk} and \ref{panel: gammaS vs q} are the same, and the maximal
efficiency occurs at a similar value of the wavelength
$\wavelength=2\pi/\wavevector$, the characteristic shapes shown in
Fig.\ \ref{gammaS vs q in the bulk} also apply to the results depicted
in Fig. \ref{panel: gammaS vs q}.

 We find that the maximal normalized swimming velocity
$\snakyEfficiency$ occurs for a similar amplitude but a shorter
wavelength $\wavelength=\wavelengthMax$ than the wavelength of a
typical $C$-shaped body posture of a nematode swimming in water [as
  depicted in Fig.\ \ref{real worms and fits}(c)].  The maximum of
$\snakyEfficiency$ occurs for $\wavelengthMax/\wormlength\approx
0.84$, whereas a typical $C$-shaped swimming body posture corresponds
to $\wavelength/\wormlength\approx 1.16$.  The wavelength at the
maximal velocity is similar to the wavelength of both the $W$-shaped
crawling gait [cf., Fig.\ \ref{real worms and fits}(a)] and gait
observed for \celegans swimming in a high-viscosity fluid.
\cite{Korta-Clark-Gabel-Mahadevan-Samuel:1007,%
  Berri-Boyle-Tassieri-Hope-Cohen:2009,%
  Fang_Yen-Wyart-Xie-Kawai-Kodger-Chen-Wen-Samuel:2010,%
  Vidal_Gadea-Topper-Young-Crisp-Kressin-Elbel-Maples-Brauner-Erbguth-Axelrod-Gottschalk-Siegel-Pierce_Shimomura:2011}
Approximate computations using HSD method yield a slightly smaller
wavelength value $\wavelengthMax/\wormlength\approx 0.70$.

Both for unconfined and confined systems the normalized swimming
velocity drops sharply for small $\wavevector\wormlength$, because the
worm body undergoes only slight deformations for
$\wavelength/\wormlength\gg 1$.  The normalized velocity also
significantly decreases at short wavelength (large
$\wavevector\wormlength$), because of finite thickness effects.  For
confined worms, there is an additional contribution to this decrease,
resulting from a smaller transverse resistance for short coherently
moving body segments (consistent with the elongated-piston mechanism
described in Sec.\ \ref{Hydrodynamic Induced Transversal and
  Longitudinal Forces}).  Therefore, the effects of finite thickness
of the nematode's body on the swimming velocity are explored in more
detail in Appendix \ref{Bead-chain model}.

The dependence of $\snakyEfficiency$ on $\amplitude/\wavevector$ is
depicted in Fig.\ \ref{gammaS vs Aq all}. For each value of
$\amplitude/\wavevector$ and channel width, the wavevector
$\wavevector\wormlength$ corresponds to the maximal value of
$\snakyEfficiency$ (cf.\ Figs. \ref{gammaS vs q in the bulk} and
\ref{panel: gammaS vs q}). Figure \ref{gammaS vs Aq all} also shows
the normalized velocity \eqref{kinematic efficiency of swimming worm}
for a worm crawling without sideways slip along the trajectories
defined by the Frenet--Serret equation \eqref{curvature diffyQ}.  We
note that even without slip, $\snakyEfficiency$ is smaller than unity,
because $\actualVelocity$ is the average velocity in the overall
direction of motion, whereas $\snakyVelocity$ is the velocity along
the curved nematode path.

Figures\ \ref{gammaS vs q in the bulk}--\ref{GammaS vs H varA} show
that confining a swimming worm in a parallel wall channel
significantly affects the normalized swimming velocity.  According to
the results plotted in Fig.\ \ref{GammaS vs H varA}, the normalized
velocity $\snakyEfficiency$ is the largest for $H/d\approx1.3$.  The peak
of $\snakyEfficiency$ is relatively broad, and for an experimentally
relevant value $H/d\approx3$, the normalized velocity is only by
25\,\% smaller than the maximal value (and twice as large as the
corresponding result for unconfined fluid, cf.\ Fig.\ \ref{gammaS vs
Aq all}. We note that enhanced swimming velocity has also been
theoretically predicted for a cylindrical swimmer in a
tube. \cite{Felderhof:2010}

The results shown in Figs.\ \ref{panel: gammaS vs q}--\ref{GammaS vs H
  varA} have been obtained using two complementary methods: the highly
accurate but numerically expensive \CRep\ method
\cite{Bhattacharya-Blawzdziewicz-Wajnryb:2005a,%
Bhattacharya-Blawzdziewicz-Wajnryb:2005} 
and HSD approximation (described in Appendix \ref{Hele--Shaw point
  dipole approximation}), which is much faster and easier to
implement.  At large wavevectors and high amplitudes of the
undulations, the HSD method overpredicts the swimming velocity.
However, it yields a good approximation for moderate values of $qL$
and $A/q$ that are relevant for swimming nematodes.  

% FIGURE 14
\begin{figure*}
\includegraphics{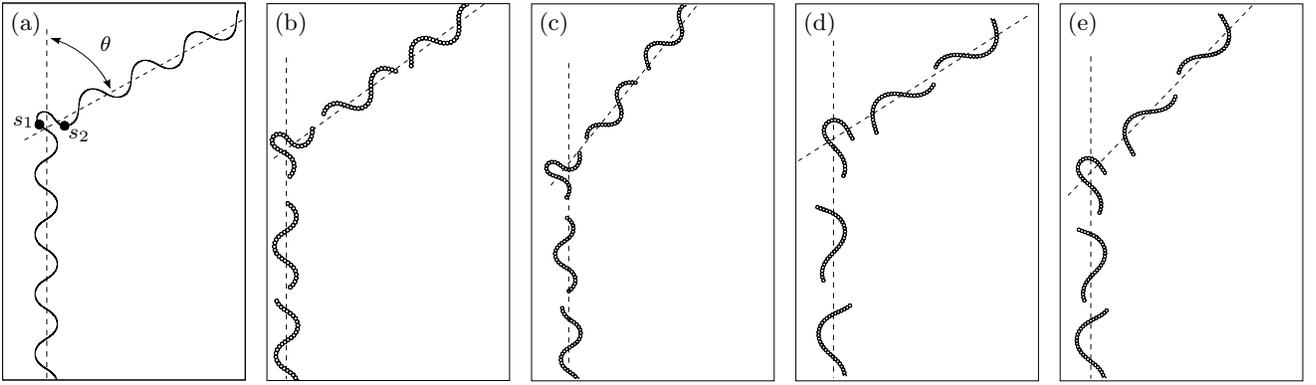}
\caption{Nematodes performing turns in different geometries: (a) worm
  crawling without slip; (b) $W$-shaped worm swimming in a
  parallel-wall channel; (c) $W$-shaped worm swimming in unconfined
  fluid; (d) $C$-shaped worm swimming in a parallel-wall channel; (e)
  $C$-shaped worm swimming in unconfined fluid.  Normalized channel
  width $H/d=1.3$.  The normalized wavevector for the $W$-shaped worm
  is $\wavevector\wormlength=9$ and for $C$-shaped worm is
  $\wavevector\wormlength=5.5$.  The turning angle and mode-switching
  points $\arcLength_1$ and $\arcLength_2$ are marked in (a); dashed
  lines indicate the direction of motion.}
\label{panel: crawling and swimming trajectories}
\end{figure*}

\section{Hydrodynamics of Turns}
\label{Hydrodynamics of Turns}

% FIGURE 15
% \input{figtex/THvS1.tex} %label= turn angle versus point of change
\begin{figure}
	\includegraphics{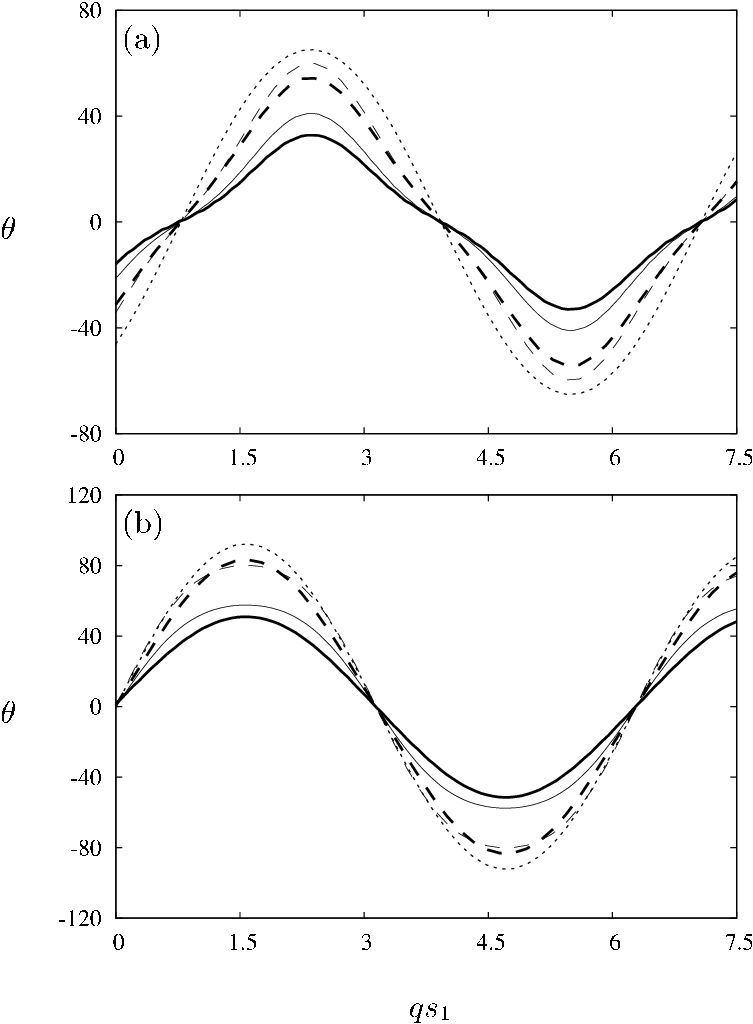}
\caption{Angle of turn $\theta$ vs.\ the normalized point of
amplitude change $\wavevector\arcLength_1$ for the length of
high-amplitude mode (a) $q\Delta s=\pi/2$ and (b) $q\Delta s=\pi$.
Crawling without slip (dotted lines).  Swimming in unconfined fluid
(solid lines) and in parallel-wall channel of width $H/d=1.3$ using
HSD (dashed lines); swimming results are presented for $C$-shaped
worms with $\wavevector\wormlength=5.5$ (thin lines) and $W$-shaped
worms with $\wavevector\wormlength=9$ (heavy lines).  The results
for a confined system evaluated using HSD approximation.}
\label{turn angle versus point of change}
\end{figure}

% FIGURE 16
% \input{figtex/THvDS.tex} %label= turn angle versus modelength
\begin{figure}
	\includegraphics{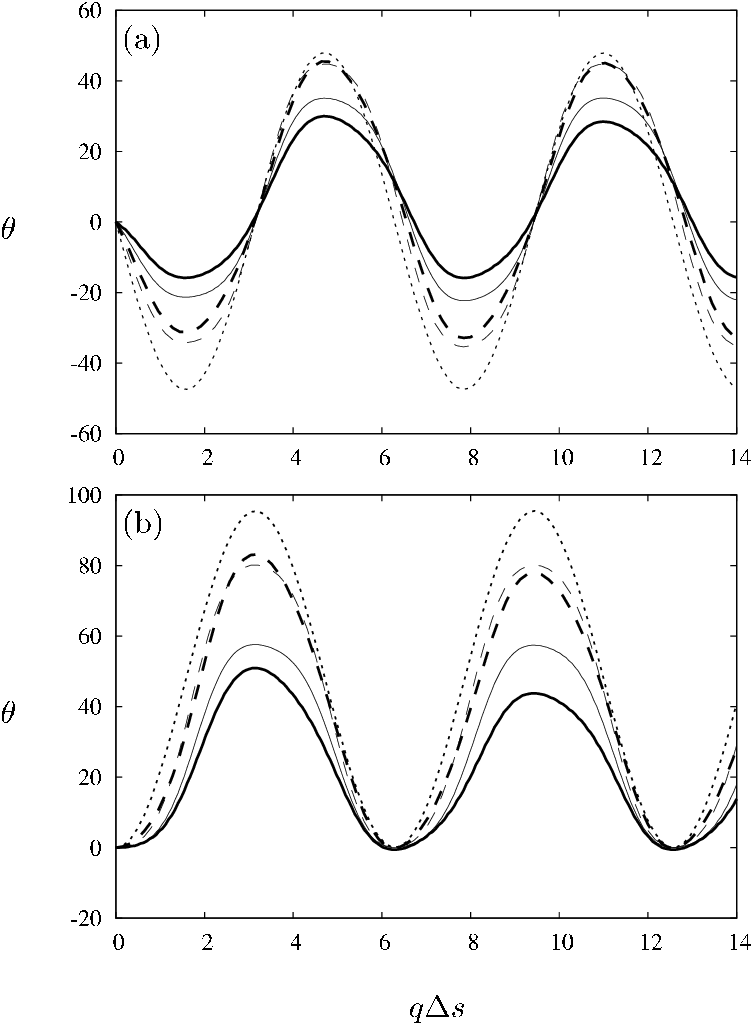}
\caption{Angle of turn $\theta$ vs.\ normalized length of
high-amplitude mode $\wavevector\Delta s$ for (a) $q s_1=0$ and (b) $q
s_1=\pi/2$; lines the same as in Fig.\ \ref{turn angle versus point of
change}.}
\label{turn angle versus modelength}
\end{figure}

Nematodes navigate their environment by performing a series of turns
to move towards the increasing concentration of a chemoattractant
\cite{Pierce-Shimomura-Dores-Lockery:2005} or in a direction of a
favorable-temperature region.\cite{Ryu-Samuel:2002}  In our recent
paper
\cite{Padmanabhan-Khan-Solomon-Armstrong-Rumbaugh-Vanapalli-Blawzdziewicz:2012}
we have shown that the nematode \celegans turns by abruptly changing
the amplitude, wavevector, and phase of the PHC function
\eqref{curvature wave}.  This behavior was thoroughly documented for a
crawling \Celegans, but our additional observations suggest that a
similar turning mechanism applies to swimming.

In this paper we consider three-mode turns, where the nematode
initially moves using a mode typical of forward locomotion, then
increases the amplitude to the $\Omega$-shaped mode
[cf.\ Fig.\ \ref{real worms and fits}(b)], and finally returns to the
initial forward-locomotion mode,
\begin{equation}
  \curvature(\arcLength) =
\label{three modes} 
    \left\{
      \begin{array}{lr}
        \amplitude_1\cos(\wavevector\arcLength),
                 \qquad  \arcLength < \arcLength_1, \\
        \amplitude_2\cos(\wavevector\arcLength),
                 \qquad \arcLength_1<\arcLength< \arcLength_2, \\
        \amplitude_1\cos(\wavevector\arcLength),
                 \qquad \arcLength_2<\arcLength.
		\end{array}
	\right.
\end{equation} 
In our calculations we use the normalized amplitude
$\amplitude_1/\wavevector=1$ for the default (forward) mode and
$\amplitude_2/\wavevector=1.8$ for the turning mode.  The normalized
wavevector $\wavevector\wormlength$ and phase $\phase$ during the turn
remain unchanged.

As illustrated in Fig.\ \ref{panel: crawling and swimming
  trajectories}(a), the turning angle of a nematode crawling without
slip depends on purely geometrical factors: the lines corresponding to
different modes \eqref{three modes} are smoothly joined together, and
the turning angle $\theta$ is obtained as a combination of the line
slopes at the joining points.\ Hence, evaluating the turning
angle only requires explicit integration of Eq.\ \eqref{curvature
  diffyQ} with respect to $\arcLength$, with the curvature given by
Eq.\ \eqref{three modes}.\ The turning angle depends on the PHC
mode parameters (including the points $\arcLength_1$ and
$\arcLength_2$ where the modes switch), but is independent of the
normalized wavevector $\wavevector\wormlength$.

For a swimming nematode, the turning angle is affected by the
rotational slip of the nematode body with respect to the surrounding
medium.  The slip depends both on the normalized amplitude
$\amplitude/\wavevector$ and the wavevector $\wavevector\wormlength$
of the curvature wave defining the sequence of shapes assumed by the
nematode.  The magnitude of the rotational slip also depends on the
confinement.  Hence, all of the above factors influence the turning
angle. \ The evolution of the worm shape is determined by
combining Eq.\ \eqref{three modes} with Eq.\ \eqref{definition of s in
  relation to time} relating the spatial variable $s$ to time.  

Turning maneuvers of unconfined and confined swimming nematodes are
illustrated in Fig.\ \ref{panel: crawling and swimming trajectories},
both for $W$-shaped and $C$-shaped nematode gaits.  As discussed in
Sec.\ \ref{efficiency}, the $W$-shaped nematode body
[Figs.\ \ref{panel: crawling and swimming trajectories}(b)
and \ref{panel: crawling and swimming trajectories}(c)] corresponds,
approximately, to the gait for which the normalized swimming velocity
$\snakyEfficiency$ assumes the maximal value.  This shape is
characteristic of nematodes crawling and swimming in high-viscosity
fluids.  The $C$-shaped body [Figs.\ \ref{panel: crawling and swimming
trajectories}(d) and \ref{panel: crawling and swimming
trajectories}(e)] corresponds to the gait observed for \celegans
swimming in water, as depicted in Fig.\ \ref{real worms and fits}(c).
The results shown in Fig.\ \ref{panel: crawling and swimming
trajectories} indicate that the same sequence of body postures that
produces a turn for a nematode crawling without slip yields a similar
turn for a swimming nematode.  However, the turning angle $\theta$ is
smaller, especially for the unconfined nematode.

Figure \ref{turn angle versus point of change} shows the turning angle
vs.\ the point of initial mode change $\arcLength_1$ for two values of
the turning-mode length $\Delta\arcLength=\arcLength_2-\arcLength_1$,
 [with $\arcLength_1=0$ corresponding to a point where
  $\curvature(s)$ has a maximum, according to Eq.\ \eqref{three
    modes}].   The dependence of $\theta$ on $\Delta\arcLength$
for two values of $\arcLength_1$ is depicted in Fig.\ \ref{turn angle
  versus modelength}.  The results indicate that the choice of mode
change parameters $\arcLength_1$ and $\Delta\arcLength$ has a
significant effect on the turning angle, both in the sign and
magnitude.  Confined swimming worms consistently perform sharper turns
than their unconfined counterparts.  The turning angles of worms
employing the $C$-shape and $W$-shape gaits are similar.  We
find that the differences between turning angles determined using the
HSD approximation and \CRep\ method (not shown) are small, with a
typical errors not exceeding $14 \%$ in the domain explored in
our simulations.

The results in Fig.\ \ref{turn angle versus modelength} are depicted
for two periods of the normalized length of the turning mode
$\wavevector\Delta\arcLength$.  For a swimming nematode, the results
in the domain $\Delta\arcLength<\wormlength$ are slightly different
than in the subsequent periods where $\Delta\arcLength>\wormlength$,
because in the first period the nematode can accommodate all three
(i.e., the initial, turning, and final) modes simultaneously along its
body.  Such three-mode body postures do not occur for the subsequent
periods of $\wavevector\Delta\arcLength$.

Since the results shown in Fig.\ \ref{turn angle versus modelength}
indicate that the effect of the three-mode postures on the turning
angles is minimal, turning angles can be accurately estimated by
combining results for single-mode and two-mode trajectories.  Such a
simplified approach significantly reduces the size of parameter space
needed to fully characterize the turning maneuvers.  In our future
study of nematode chemotaxis, this simplified method will increase the
efficiency of numerical simulations.

\section{Conclusions}
\label{conclusions}

Combining our PHC description of the nematode gait
\cite{%
Padmanabhan-Khan-Solomon-Armstrong-Rumbaugh-Vanapalli-Blawzdziewicz:2012} 
and highly accurate hydrodynamic models, we have quantitatively
characterized locomotion capabilities of swimming submillimeter-size
nematodes.  We have investigated the swimming velocity and turning
maneuvers for locomotion in unconfined fluid and in a fluid confined
by two parallel walls.  The swimming effectiveness was characterized
by the swimming velocity $\snakyEfficiency$ normalized by the velocity
of curvature wave propagating along the nematode body.

We have determined the dependence of the normalized velocity
$\snakyEfficiency$ on the wavevector $\wavevector$ and amplitude
$\amplitude$ of the curvature wave.  It has been found that the
velocity is maximal for the normalized amplitude
$\amplitude/\wavevector\approx1$, consistent with our earlier
experimental study of the gait of a swimming nematode.
\cite{%
Padmanabhan-Khan-Solomon-Armstrong-Rumbaugh-Vanapalli-Blawzdziewicz:2012}
However, the wavelength $\wavelengthMax$ of the gait corresponding to
the maximal velocity is shorter than the experimentally observed
wavelength for nematodes swimming in water.  We determined that the
calculated wavelength is similar to the one characteristic of a
nematode swimming in a high-viscosity fluid.
\cite{Fang_Yen-Wyart-Xie-Kawai-Kodger-Chen-Wen-Samuel:2010}
In a forthcoming publication we will show that the choice of swimming
gait can be explained using energy-dissipation considerations: the
gait change stems from a different wavelength dependence of the
hydrodynamic energy dissipation in the external fluid and the internal
dissipation in the nematode body.

Our numerical simulations of nematodes swimming in a parallel-wall
channel reveal that confinement can significantly enhance the swimming
velocity.  This behavior stems from the increased transverse
hydrodynamic resistance due to a large pressure drop across the
nematode body moving sideways in a narrow space.  We find that for the
confinement ratio $H/d=3$ the normalized swimming velocity
$\snakyEfficiency$ exceeds by a factor of two the swimming velocity of
an unconfined nematode.  The effect of the increased swimming velocity
should be taken into account in interpretation of experimental
observations of nematode locomotion in parallel-plate cells.

The enhanced swimming velocity under strong confinement in a
parallel-wall cell was, in fact, observed in recent experiments.
\cite{Lebois-Sauvage-Py-Cardoso-Ladoux-Hersen-Di_Meglio:2012} Our
results provide the explanation of this phenomenon.   An
increased locomotion efficiency was also seen for \Celegans\ moving in
microfabricated pillar environments.
\cite{Majmudar-Keaveny-Zhang-Shelley:2012,%
Park-Hwang-Nam-Martinez-Austin-Ryu:2008} 
However, in a pillar system \Celegans\ produces effective propulsion
by pushing against mechanical obstacles, whereas the effect described
in our study is of purely hydrodynamic origin.

The analysis of turning maneuvers shows that turns in swimming and
turns in crawling can be performed using the same set of body
postures.  The turning angle is larger for a worm crawling without
slip, but the angles in swimming are sufficiently large for effective
maneuverability.  To our knowledge, this study is the first systematic
hydrodynamic investigation of turning maneuvers in undulatory
locomotion.  Results of our hydrodynamic calculations of swimming
speed and turning angles have immediate applications in modeling
chemotaxis of nematodes immersed in water,
\cite{Patel-Bilbao-Padmanabhan-Khan-Armstrong-Rumbaugh-Vanapalli-Blawzdziewicz:2012}
thermotaxis and electrotaxis.  We are also using these results to
study motor control of a swimming \Celegans.

\begin{acknowledgments}
We would like to acknowledge financial support from NSF Grant
No.\ CBET 1059745 (A.\,B.\ and J.\,B.) and National Science Center
(Poland) Grant No.\ 2012/05/B/ST8/03010 (E.\,W.).
S.\,A.\,V.\ acknowledges NSF CAREER Award Grant No.\ 1150836.

\end{acknowledgments}

\appendix

\section{Active bead-chain model}
\label{Bead-chain model}

In our calculations the body of the nematode is modeled as a long
active chain of $\lengthofchain$ touching beads.  The chain undergoes
deformations  that mimic undulatory motion of a
swimming worm (cf.\ Fig.\ \ref{bead model}).  The overall
translational and rotational motion of the chain results from the
balance of the hydrodynamic forces and torques acting on the beads.
Chain kinematics is described in Sec.\ \ref{Chain kinematics}, and
chain hydrodynamics is analyzed in Sec.\ \ref{hydrodynamic
  interactions}.

\subsection{Chain kinematics}
\label{Chain kinematics}

Consistent with the description of nematode kinematics introduced in
Secs.\ \ref{kinematics} and \ref{Force Balance} (see
Fig.\ \ref{snapshot sequence crawling versus swimming}), the beads
move along the line $\lineS$ defined by the Frenet--Serret equations
\eqref{curvature diffyQ} with the piecewise-harmonic curvature
\eqref{harmonic curvature multiple}.  In the lab coordinate system
$\labCoordinateSystem$, the line $\lineS$ rotates and translates with
the linear and angular velocities $\velocityVecRB$ and
$\rotvelocityVecRB$.  Accordingly, the translational and rotational
velocities of each bead,
\begin{subequations}
\label{total linear and angular bead velocities}
\begin{eqnarray}
  \velocityVeci{i}&=&\velocityVecSi{i}+\velocityVecRBi{i},\\
\label{total velocity of a bead}
  \rotvelocityVeci{i}&=&\rotvelocityVecSi{i}+\rotvelocityVecRBi{i},
\label{total angular velocity of a bead}
\end{eqnarray}
\end{subequations}
have the active components $\velocityVecSi{i}$ and
$\rotvelocityVecSi{i}$ associated with the forward motion along the
line $\lineS$, and rigid-body components
\begin{subequations}
\label{rigid body translation and rotation per bead}
\begin{eqnarray}
\velocityVecRBi{i}&=&\velocityVecRB
   +\rotvelocityVecRB\vecproduct\distancei{i},\\
\label{rigid body velocity definition per bead}
\rotvelocityVecRBi{i}&=&\rotvelocityVecRB,
\label{rigid body rotation definition per bead}
\end{eqnarray}
\end{subequations}
where $\distancei{i}$ is the position of the bead $i$.

The active component of the linear velocity of bead $i$ is given by
the relation
\begin{equation}
  \velocityVecSi{i}=\snakyVelocity \unittangentVectori{i},
\label{velocity per bead}
\end{equation}
where $\unittangentVectori{i}$ is the unit vector tangent to the curve
$\lineS$ at the position of bead $i$.  Relation \eqref{velocity per
  bead} corresponds to the active rod moving along the line $\lineS$
with velocity $\snakyVelocity$ and is fully determined by the system
geometry.  In contrast, the angular velocities of the beads
\begin{equation}
  \rotvelocityVecSi{i}=\rotvelocitySi{i}\ez
\label{angular velocity per bead}
\end{equation}
(where $\ez$ is the unit vector in the direction $z$ normal to the
plane of motion), are not uniquely determined, except for a chain
moving along a line with constant curvature
$\curvature(\arcLength)=\curvature_0$.  In this case the angular
velocities of all beads are the same,
\begin{equation}
\label{angular velocities for circular motion}
\omega_i=\kappa_0\snakyVelocity,
\end{equation}
because the chain moves along the circle as a rigid body.  [For a
  force-free and torque-free chain submerged in a fluid the velocity
  component \eqref{rigid body translation and rotation per bead}
  compensates for the imposed motion \eqref{angular velocity per bead}
  and \eqref{angular velocities for circular motion}, so that a
  circular chain remains at rest.]  To verify the accuracy of our
approach, we examine three internally consistent bead-rotation models
for a flexible chain moving along a line with a varying curvature
$\curvature(\arcLength)$.  We find that at short wavelengths bead
rotations have a large effect on the motion of a chain, but the choice
of a specific rotation model does not influence the results in a
significant way.

\paragraph*{Local-curvature model.---} In this model individual bead 
rotations are evaluated according to the local formula
\begin{equation}
\rotvelocitySi{i}
=
\kappa(\arcLength_i)\snakyVelocity,
\label{simple bead rotation model}
\end{equation}
where $\arcLength_i$ is the value of the coordinate $\arcLength$ for
bead $i$.  The model \eqref{simple bead rotation model} satisfies the
consistency condition \eqref{angular velocities for circular motion}, but it introduces a relative slip of particle surfaces.

\paragraph*{Model with no interparticle slip.---} Bead angular 
velocities $\rotvelocitySi{i}$ are chosen in such a way that
interparticle slip is not present.  Accordingly, it is assumed that
the angular velocities of the beads satisfy the no-relative-slip
condition
\begin{equation}
\rotvelocitySi{i}+\rotvelocitySi{i+1}=2 \rodrotvelocityi{i,i+1},\qquad
i=1,2\ldots \lengthofchain-1,
\label{individual bead rotation noslip}
\end{equation}
where $\rodrotvelocityi{i,i+1}$ is the angular velocity of the vector
connecting beads $i$ and $i+1$, which is computed as
\begin{equation}
\rodrotvelocityi{i,i+1}
=
\frac{d\rodbetbeadsi{i,i+1}}{dt}
\dotproduct
\unitnormalVectori{i,i+1}.
\label{definition of rod rotational velocity noslip}
\end{equation}
Here $\rodbetbeadsi{i,i+1}$ denotes the unit vector along the line
connecting beads $i$ and $i+1$, and $\unitnormalVectori{i,i+1}$ is the
unit vector normal to $\rodbetbeadsi{i,i+1}$. The system of equations
\eqref{individual bead rotation noslip} is solved subject to the
boundary condition
\begin{equation}
\rotvelocitySi{\lengthofchain}
=
\kappa(s_\lengthofchain)\snakyVelocity,
\label{boundary condition noslip}
\end{equation}
which ensures the consistency condition \eqref{angular velocities for
  circular motion}.

\paragraph*{Model with smoothed angular velocity.---} 
This approach aims to more accurately mimic the motion of deformable
interface of the nematode for a system with a strongly nonlinear
variation of the curvature.  To smooth out the effect of rapid
curvature variation, the angular velocity of bead $i$ is given as the
average rate of rotation of the directors $\rodbetbeadsi{i-1,i}$ and
$\rodbetbeadsi{i,i+1}$.  Accordingly, the bead angular velocities are
given by
\begin{equation}
\rotvelocitySi{i}=\textstyle\frac{1}{2}
(\rodrotvelocityi{i-1,i}+\rodrotvelocityi{i,i+1}),\qquad i=2,3 \ldots N-1.
\label{individual bead rotation rod method}
\end{equation}
The angular velocities of the first and  last beads are
\begin{equation}
\rotvelocitySi{1}=\rodrotvelocityi{1,2},\qquad
    \rotvelocitySi{N}=\rodrotvelocityi{\lengthofchain-1,\lengthofchain}.
\label{tail boundary condition rods}
\end{equation}

% FIGURE 17
% \input{figtex/GammaVSq_varM_bulk.tex} %label= gammaS vs q in the bulk for varM
\begin{figure}
	\includegraphics{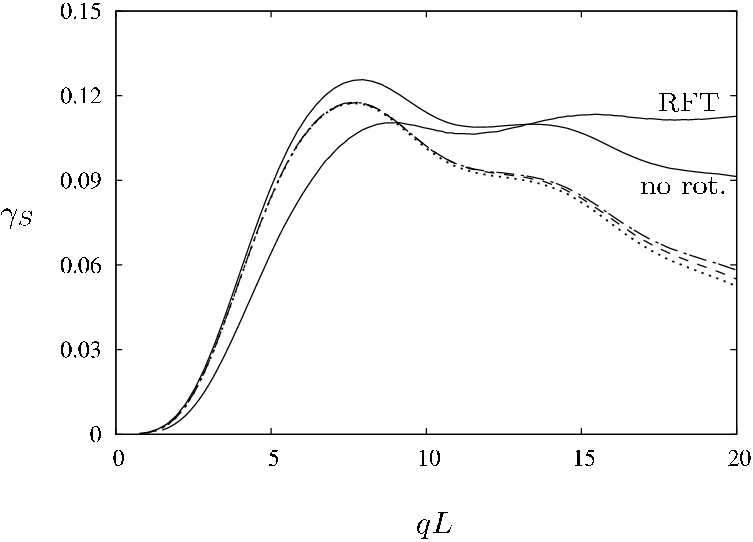}
\caption{ Normalized swimming velocity $\snakyEfficiency$ vs. wave
vector $\wavevector$ normalized by the worm length $\wormlength$ for
a nematode swimming in unconfined fluid.  Local-curvature model
(dashed dotted line), model with smoothed angular velocity (dashed
line) and model with no interparticle slip (dotted line). Solid
lines represent resistive force theory and no rotation model as
labeled. }
\label{gammaS vs q in the bulk for varM}
\end{figure}

The effect of bead rotation on the swimming speed of an active rod is
illustrated in Fig.\ \ref{gammaS vs q in the bulk for varM}.  By
comparing the results that include consistent bead-rotation models
with a calculation that neglects bead rotations entirely, we find that
the angular velocities of the beads have a significant effect on the
chain velocity.  However, if the bead rotations are properly
implemented, discrepancies between various rotation models are small,
with appreciable differences occurring only in the regime where the
wavevector normalized by the bead diameter is too large, $q d\gtrsim
O(1)$.  In this regime noticeable differences are expected even for a
continuous elongated body, because deformation details are not
uniquely determined by the curvature of centerline alone.

In Fig.\ \ref{gammaS vs q in the bulk for varM} the results of the
bead-chain model are also compared to the normalized swimming velocity
evaluated using the RFT with the resistance-coefficient ratio
$\transforce/\longforce= 1.45$ (corresponding to a coherently moving
15-bead chain segment).  The results indicate that RFT does not
capture the decrease of the swimming velocity at short waves, and
therefore significantly overpredicts the normalized swimming velocity
in this regime.

\subsection{Hydrodynamic interactions}
\label{hydrodynamic interactions}

Under creeping-flow conditions, the hydrodynamic force and torque
acting on bead $i$ in a chain moving through a viscous fluid is
related to linear and angular bead velocities via the
$\lengthofchain$-particle friction relation
\begin{equation}
\begin{bmatrix}
    \forcei{i} \\ 
    \torquei{i} 
\end{bmatrix}
      =\sum_{j=1}^\lengthofchain
\begin{bmatrix}
    \Ztt_{ij} & \Ztr_{ij} \\
    \Zrt_{ij} & \Zrr_{ij}
\end{bmatrix}
\dotproduct
\begin{bmatrix}
    \velocityVeci{j} \\ 
    \rotvelocityVeci{j}
\end{bmatrix},
\label{torque force relation for chain of spheres}
\end{equation}
where $\Z^{\alpha\beta}_{ij}$ ($\alpha,\beta=t,r$) are the
translational ($t$) and rotational ($r$) hydrodynamic resistance
coefficients.

The total force acting on the chain is equal to the sum of the
forces on individual beads
\begin{equation}
  \totalForce=\sum_{i=1}^\lengthofchain \forcei{i}.
  \label{sum for forces}
\end{equation}
The total torque
\begin{equation}
\totalTorque=\sum_{i=1}^\lengthofchain
  \left(
    \torquei{i}+\distancei{i}\vecproduct\forcei{i} 
  \right)
\label{sum for torques}
\end{equation}
includes the sum of individual bead torques $\torquei{i}$ as well as
the torque due to the forces acting on the beads.  Taking into account
the velocity decomposition \eqref{total linear and angular bead
velocities}, the total force and torque can be represented as a
superposition of the active and hydrodynamic-resistance components,
\begin{equation}
\label{total force and torque decomposed into active and passive
components}
\begin{bmatrix}
    \totalForce \\ 
    \totalTorque 
\end{bmatrix}
  =
  \begin{bmatrix}
    \totalForceActive \\ 
    \totalTorqueActive
  \end{bmatrix}
  +
  \begin{bmatrix}
    \totalForceHR \\ 
    \totalTorqueHR
  \end{bmatrix},
\end{equation}
where
\begin{equation}
\label{active force and torque}
  \begin{bmatrix}
    \totalForceActive \\ 
    \totalTorqueActive
  \end{bmatrix}
    =
\begin{bmatrix} 
  \Zts \\ 
  \Zrs  
\end{bmatrix}
    \snakyVelocity
\end{equation}
and
\begin{equation}
\label{resistance force and torque}
  \begin{bmatrix}
    \totalForceHR \\ 
    \totalTorqueHR
  \end{bmatrix}
    =
\begin{bmatrix}
  \Ztt & \Ztr \\
  \Zrt & \Zrr
\end{bmatrix}
  \dotproduct
\begin{bmatrix} 
  \velocityVecRB \\
  \rotvelocityVecRB
\end{bmatrix}.
\end{equation}
Combining expressions \eqref{torque force relation for chain of
spheres}--\eqref{sum for torques} with the bead rotation models
of Sec. \ref{Chain kinematics} we find the active-force
matrix
\begin{equation}
\label{active force and torque coefficients}
\begin{bmatrix} 
  \Zts \\ 
  \Zrs  
\end{bmatrix}
    =\sum_{i,j=1}^\lengthofchain
  \begin{bmatrix}
     \identity&0\\
     \Rcrossi&\identity
  \end{bmatrix}
    \dotproduct
  \begin{bmatrix}
    \Ztt_{ij} & \Ztr_{ij} \\
    \Zrt_{ij} & \Zrr_{ij}
  \end{bmatrix}
    \dotproduct
  \begin{bmatrix}
    \barVelocityVecSi{j}\\
    \barRotvelocityVecSi{j}
  \end{bmatrix},
\end{equation}
(where $\barVelocityVecSi{j}= \velocityVecSi{j}/\snakyVelocity$ and
$\barRotvelocityVecSi{j}=\rotvelocityVecSi{j}/\snakyVelocity$ are the
normalized active linear and angular velocities of the particles), and
the chain-resistance hydrodynamic matrix
\begin{equation}
\label{chain friction tensor}
  \begin{bmatrix}
    \Ztt & \Ztr \\
    \Zrt & \Zrr
  \end{bmatrix}
    =\sum_{i,j=1}^\lengthofchain
  \begin{bmatrix}
     \identity&0\\
     \Rcrossi&\identity
  \end{bmatrix}
    \dotproduct
  \begin{bmatrix}
    \Ztt_{ij} & \Ztr_{ij} \\
    \Zrt_{ij} & \Zrr_{ij}
  \end{bmatrix}
    \dotproduct
  \begin{bmatrix}
     \identity&\Rcrossj^{\dagger}\\
     0&\identity
  \end{bmatrix}.
\end{equation}
In the above equations, $\identity$ is the identity tensor, the dagger
denotes the transpose, and we use the cross-product operator notation
\begin{equation}
    \Rcrossi \dotproduct {\bf A} \\
  =
    \distancei{i} \vecproduct {\bf A}\\
\end{equation}
where ${\bf A}$ is an arbitrary vector.  Equations
\eqref{active force and torque coefficients} and \eqref{chain friction
  tensor} provide a link between the active bead-chain model and the
hydrodynamic description of nematode locomotion given in
Eqs.\ \eqref{total force of a body}--\eqref{mobility matrix
  definition}.

For a bead chain moving in an unconfined space, the multiparticle
hydrodynamic resistance matrix $\Z^{\alpha\beta}_{ij}$ is evaluated
using the {\sc Hydromultipole} method.
\cite{Cichocki-Felderhof-Hinsen-Wajnryb-Blawzdziewicz:1994} For a
parallel-wall geometry we use the 
{\CRep} method.
\cite{Bhattacharya-Blawzdziewicz-Wajnryb:2005a,%
Bhattacharya-Blawzdziewicz-Wajnryb:2005}
In both cases, the lubrication resistance for touching neighboring
beads is truncated at a finite gapwidth $\epsilon_0$ between particle
surfaces.  This truncation allows us to avoid infinite internal
friction in the chain.  The active force and resistance coefficients
\eqref{active force and torque} and \eqref{chain friction tensor} are
non-singular in the limit $\epsilon_0\to0$.  We find that the
numerical results for chain motion  are insensitive to the value of
$\epsilon_0$ for $\epsilon_0\ll1$.

 Evaluation of the hydrodynamic-interaction tensors using the
       {\CRep} method is very accurate but numerically expensive.
       Thus we have also developed a less accurate but much faster
       Hele--Shaw dipole (HSD) approximation, described in Appendix
       \ref{Hele--Shaw point dipole approximation}.  A comparison of
       HSD results with the CR method, presented in Fig.\ \ref{panel:
         gammaS vs q}, shows that the HSD approximation provides
       accurate description of the hydrodynamics of active bead chains
       at sufficiently long wavelengths.  

\begin{figure}[b]
\includegraphics{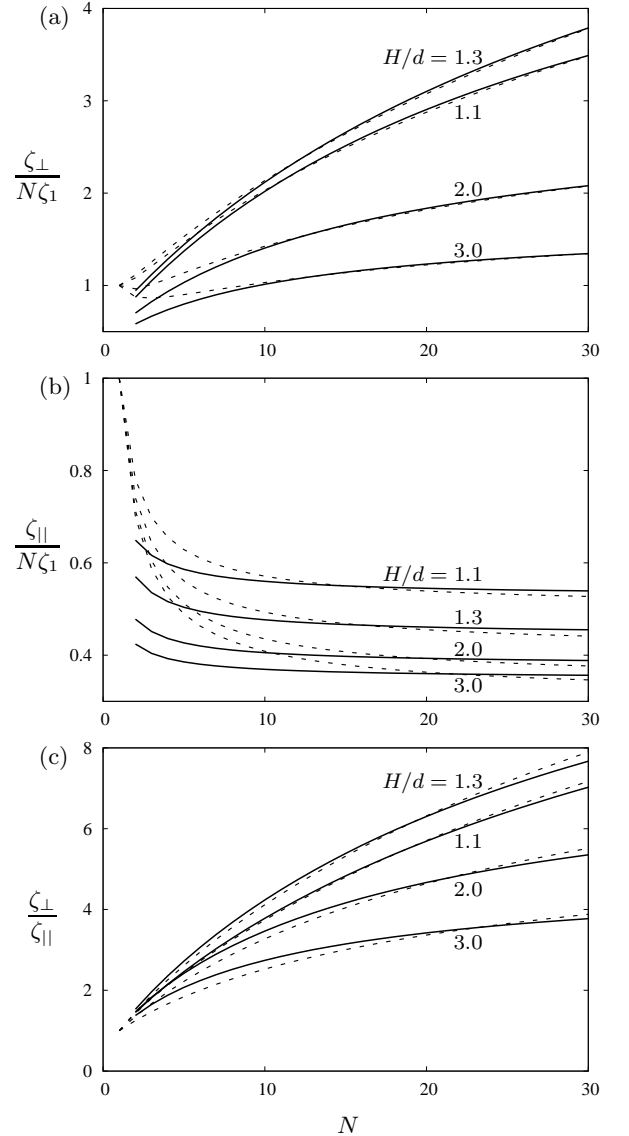}
\caption{A comparison of the HSD approximation (solid lines) with
accurate results obtained using {\CRep} method (dashed lines) for
the transverse and longitudinal resistance coefficients of linear
chains of touching spheres in the midplane of a parallel-wall
channel.  The (a) transverse and (b) longitudinal resistance
coefficients per particle (normalized by the one-particle value) and
(c) the resistance-coefficient ratio are shown vs.\ the chain length
$\lengthofchain$ for the normalized channel width $H/d$, as labeled.}
\label{hele shaw fitting plot}
\end{figure}

\section{Hele--Shaw dipole approximation}
\label{Hele--Shaw point dipole approximation}

\subsection{Interparticle dipolar interactions}
\label{hele-shaw point dipole approximation}

An isolated spherical particle $i$ in the midplane of a parallel-wall
channel, moving with the velocity $\velocityVeci{i}$ and subject to
the external pressure gradient $\bnabla\externalPressure$ experiences
the hydrodynamic traction force $\forcei{i}$ and produces the
far-field scattered parabolic flow driven by a two-dimensional pressure
dipole,
\cite{Cui-Diamant-Lin-Rice:2004,%
Bhattacharya-Blawzdziewicz-Wajnryb:2005,%
Bhattacharya-Blawzdziewicz-Wajnryb:2006,%
Blawzdziewicz-Wajnryb:2008,%
Janssen-Baron-Anderson-Blawzdziewicz-Loewenberg-Wajnryb:2012%
}
\begin{equation}
\scatteredPressure(\lateralPositionVectori{i}')
  =\dipolarMomenti{i}\dotproduct
   \frac{\lateralPositionVectori{i}'}{{\lateralPositioni{i}'}^2}.
\label{pressure produced by a particle}
\end{equation}
Here
$\lateralPositionVectori{i}'=\lateralPositionVector-\lateralPositionVectori{i}$
is the lateral position of the field point
$\lateralPositionVector=x\ex+y\ey$ relative to the lateral particle
position $\lateralPositionVectori{i}=X_i\ex+Y_i\ey$,
$\lateralPosition=|\lateralPositionVector|$, and $\dipolarMomenti{i}$
is the dipolar moment of the induced pressure dipole.  The
hydrodynamic traction force and dipolar moment are linearly related to
the particle velocity $\velocityVeci{i}$ and the external pressure
gradient at the particle position,
\begin{equation}
\label{pressure gradient at particle position}
\bnabla\externalPressurei{i}
   \equiv\bnabla\externalPressure(\lateralPositionVectori{i})
\end{equation}
(where it is assumed that $\externalPressure$ depends only on the
lateral coordinates).  As discussed in Ref.\
\onlinecite{Blawzdziewicz-Wajnryb:2008},
the force and dipolar moment can be expressed by the generalized
resistance relation
\begin{equation}
  \begin{bmatrix}
    \forcei{i}\\
    \dipolarMomenti{i}
  \end{bmatrix}
    = 
  \begin{bmatrix}
    \zetaHSD^{tt} & \zetaHSD^{tp} \\
    \zetaHSD^{pt} & \zetaHSD^{pp}
  \end{bmatrix}
    \dotproduct
  \begin{bmatrix}
    \velocityVeci{i}\\
    \bnabla\externalPressurei{i}
  \end{bmatrix},
\label{force and dipole related to velocity and pressure}
\end{equation}
where the scalar resistance coefficients $\zetaHSD^{\alpha\beta}$
depend on the confinement ratio $H/d$.

In the HSD approximation it is assumed that the particles interact
solely via the Hele-Shaw dipolar fields \eqref{pressure produced by a
particle}.  It follows that the flow incident to particle $i$ is
driven by the pressure gradient
\begin{equation}
  \bnabla \externalPressurei{i} 
    =\bnabla\sum_{j \neq i}^\lengthofchain
      \scatteredPressure(\lateralPositionVectori{j}')
\label{external pressure field}
\end{equation}
resulting from the superposition of dipolar pressures \eqref{pressure
  produced by a particle} produced by other particles.  By combining
Eqs.\ \eqref{force and dipole related to velocity and pressure} and
\eqref{external pressure field}, we obtain the relations
\begin{subequations}
\label{Hele-Shaw dipole equations}
\begin{eqnarray}
\forcei{i}&=&\zetaHSD^{tt} \velocityVeci{i}
           +\zetaHSD^{tp}\eta^{-1}\sum_{j \ne i}^\lengthofchain 
            \positiontensor _{ij}\dotproduct\dipolarMomenti{j},\\
\label{hele-shaw force vector}
  \dipolarMomenti{i}&=&\zetaHSD^{pt}\velocityVeci{i}
                   +\zetaHSD^{pp}\eta^{-1}\sum_{j \ne i}^\lengthofchain 
                    \positiontensor_{ij}
			\dotproduct
			\dipolarMomenti{j},
\label{hele-shaw dipole vector}
\end{eqnarray}
\end{subequations}
where
\begin{equation}
\positiontensor_{ij}
  =\frac
    {\identity-2\lateralPositionVectori{ij}\lateralPositionVectori{ij}}
    {\lateralPositioni{ij}^2}
\label{definition of g tensor}
\end{equation}
is the dipolar-interaction tensor obtained by taking the gradient of
Eq.\ \eqref{pressure produced by a particle}.

Eliminating the dipole moment $\dipolarMomenti{i}$ from equations
\eqref{Hele-Shaw dipole equations} yields the
$\lengthofchain$-particle friction relation 
\begin{equation}
\forcei{i}=\sum_{i,j=1}^\lengthofchain\ZHStti{ij}\dotproduct\velocityVeci{j}
\label{hele-shaw force relation final}
\end{equation}
in the HSD approximation. The hydrodynamic resistance tensor
$\ZHStti{ij}$ can be expressed using the $\lengthofchain$-particle
matrix relation,
\begin{equation}
\ZHStt=\zetaHSD^{tt}\identityN+\eta^{-1}\zetaHSD^{tp}\zetaHSD^{pt} 
  \left[
   \identityN-\eta^{-1}\zetaHSD^{pp}\positiontensor
\right]^{-1},
\label{definition of hele-shaw friction matrix}
\end{equation}
where $\ZHStt$ and $\positiontensor$ are
$\lengthofchain\times\lengthofchain$ matrices with elements
$\ZHStti{ij}$ and $\positiontensor_{ij}$
($i,j=1,\ldots,\lengthofchain$), and $\identityN$ is the identity
matrix in the $\lengthofchain$-particle space.

In the HSD approximation, the dynamics of the active chain is
described by the active-force and chain-resistance relations
\eqref{active force and torque coefficients} and \eqref{chain friction
  tensor} with
\begin{subequations}
\label{resistance matrix in HSD approximation}
\begin{equation}
\label{translational resistance in HSD approximation}
\Ztt_{ij}=\ZHStti{ij}
\end{equation}
for the translational components of the $\lengthofchain$-particle
hydrodynamic resistance matrix, and
\begin{equation}
\label{rotational resistance components in HSD approximation}
\Ztr_{ij}=\Zrt_{ij}=\Zrr_{ij}=0
\end{equation}
\end{subequations}
for the components that involve particle rotation.  The rotational
components \eqref{rotational resistance components in HSD
  approximation} are neglected in the HSD approximation, because
particle rotation in the midplane of the channel does not produce
far-field dipolar scattered flow.\cite{Blawzdziewicz-Wajnryb:2008}
However, we expect that incorporating short-range rotational effects
in future implementations of the HSD method may improve its accuracy
at short wavelengths.

\subsection{Application to a system of touching spheres}
\label{Numerical Methods}

According to Eq.\ \eqref{definition of hele-shaw friction matrix}, the
HSD approximation involves three independent numerical coefficients,
i.e., $\zetaHSD^{tt}$, $\zetaHSD^{pp}$, and the product
$\zetaHSD^{tp}\zetaHSD^{pt}$, which describe the single-particle
hydrodynamic response \eqref{force and dipole related to velocity and
  pressure} of a particle to its translational motion and to the
applied pressure gradient. When the values corresponding to the
dynamics of an isolated particle are used for these coefficients, an
asymptotic approximation for a system of widely separated particles is
obtained.  Such an approximation, however, is inaccurate if the
interparticle distance is comparable to the channel width.
\cite{Baron-Blawzdziewicz-Wajnryb:2008}

In the present study the HSD approximation is applied outside the
far-field asymptotic regime. Therefore, we use an alternative
approach, where we treat the coefficients $\zetaHSD^{tt}$,
$\zetaHSD^{pt}$, and $\zetaHSD^{tp}\zetaHSD^{pt}$ as adjustable
parameters.  The values of these parameters are determined by fitting
the HSD results to accurate CR calculations for the resistance
coefficients of rigid linear chains of touching spheres.  The same
values are then used in our simulations of the motion of active
particle chains.

For a given normalized channel width $H/d$, the optimal parameter
values are determined by minimizing the fitting error
\begin{widetext}
\begin{equation}
f=\sum_{i=\Nmin}^{\Nmax}\left\{
  \alpha\left[\longforceExact{i}-\longforceHSD{i}\right]^2
    +
        \left[\transforceExact{i}-\transforceHSD{i}\right]^2 
                        \right\}^{1/2},
\label{function to be minimized}
\end{equation}
\end{widetext}
where $\longforce$ and $\transforce$ are the lateral and transverse
resistance coefficients of chains of different lengths $i$.  The
indices $\exact$ and $\HS$ refer to the results obtained using the
{\CRep} algorithm
\cite{Bhattacharya-Blawzdziewicz-Wajnryb:2005a,%
Bhattacharya-Blawzdziewicz-Wajnryb:2005}
and HSD approximation \eqref{definition of hele-shaw friction matrix},
respectively.

\begin{table}[position specifier]
	\begin{tabular}{p{1cm} p{1.5cm} p{1.5cm} p{1cm}}
		\hline
		\hline
		$H/d$ & $\zetaHSD^{tt}$ & $\zetaHSD^{tp}\zetaHSD^{pt}$ & $\zetaHSD^{pp}$ \\
		\hline
 		1.01 & 0.85223 & 0.05176 & 0.15010 \\
 		1.02 & 0.82923 & 0.05748 & 0.15001 \\
 		1.03 & 0.80829 & 0.06030 & 0.15032 \\
 		1.04 & 0.79819 & 0.06347 & 0.15003 \\
 		1.05 & 0.78857 & 0.06559 & 0.14990 \\
 		1.06 & 0.78352 & 0.06726 & 0.14976 \\
 		1.07 & 0.78188 & 0.06910 & 0.14945 \\
 		1.08 & 0.76290 & 0.06866 & 0.14994 \\
 		1.09 & 0.76152 & 0.07058 & 0.14948 \\
		1.1 & 0.75416 & 0.07068 & 0.14960 \\
		1.2 & 0.70692 & 0739421 & 0.14838 \\
		1.3 & 0.67887 & 0.07304 & 0.14705 \\
		1.4 & 0.65012 & 0.07020 & 0.14570 \\
		1.5 & 0.64069 & 0.06973 & 0.14323 \\
		2.0 & 0.56027 & 0.05441 & 0.13770 \\
		2.5 & 0.52446 & 0.04719 & 0.13036 \\
		3.0 & 0.48439 & 0.03928 & 0.12826 \\
		\hline
	\end{tabular}
\caption{Coefficients $\zetaHSD^{tt}$, $\zetaHSD^{pp}$, and
  $\zetaHSD^{tp}\zetaHSD^{pt}$ of the Hele--Shaw dipole approximation
  \eqref{definition of hele-shaw friction matrix} for active chains of
  touching spheres, for different values of dimensionless channel
  width.}
\label{table relating height with zeta parameters}
\end{table}

In our calculations we have used the summation range from $\Nmin=3H/d$
to $\Nmax=30$ and the value $\alpha=4$ for the weight of the lateral
resistance component relative to the transverse component.  The lower
summation limit $\Nmin$ corresponds to the chain length below which
the HSD approximation is not expected to hold, based on the decay
distance of the near-field contributions.
\cite{Bhattacharya-Blawzdziewicz-Wajnryb:2006,%
Baron-Blawzdziewicz-Wajnryb:2008}
The upper limit $\Nmax$ is the chain length used in our simulation of
nematode locomotion.  The fitting error \eqref{function to be
  minimized} was minimized using the conjugate-gradient method.

The results of our calculations are illustrated in Fig.\ \ref{hele
  shaw fitting plot}, where we compare HSD approximation with accurate
results obtained using the {\CRep} method.  For sufficiently large
chain lengths $\lengthofchain$, the HSD approximation agrees well with
the accurate calculations, especially for narrow channels.  The error
for the resistance-coefficient ratio $\forceratio$ is small in the
whole range of chain lengths $\lengthofchain$, as depicted in
Fig.\ \ref{hele shaw fitting plot}(c).  The optimal values for the
coefficients $\zetaHSD^{tt}$, $\zetaHSD^{pp}$, and the product
$\zetaHSD^{tp}\zetaHSD^{pt}$ for different channel widths are listed
in Table \ref{table relating height with zeta parameters}.

The above fitting procedure optimizes the accuracy of the resistance
coefficients for linear chains of spheres.  Since the motion of
deformable active chains is determined by hydrodynamic interactions of
coherently moving chain segments, the HSD approximation yields
accurate results for active chains at sufficiently long wavelengths.

%\bibliography{/home/albilbao/BIB/albib,\wherebibJ/jbib,\wherebibJ/vbibx,\wherebibJ/elegans,\wherebibJ/zymurgy}

%\bibliography{\wherebibJ/albib,\wherebibJ/jbib,\wherebibJ/vbibx,\wherebibJ/elegans,\wherebibJ/zymurgy}

\bibliography{\wherebibJ/jbib,\wherebibJ/vbibx,\wherebibJ/elegans,\wherebibJ/zymurgy}

\end{document}